\documentclass[preprint,eqsecnum,preprintnumbers,nofootinbib,byrevtex,prd,aps,showpacs,showkeys,groupedaddress,floatfix]{revtex4}
\usepackage{bm}
\usepackage{graphics}
\usepackage{graphicx}
\usepackage{epsfig}
\usepackage{amssymb}
\usepackage{amsmath}
\begin{document}
\large
\title{INVESTIGATION OF THE NEUTRALINO PAIR PRODUCTION AT LHC}
\author{A.~I.~Ahmadov$^{1,2}$}
\email{E-mail:ahmadovazar@yahoo.com}
\author{I.~Boztosun$^{1}$}
\author{R.~Kh.~Muradov$^{2}$}%
\author{A.~Soylu$^{1}$}
\author{E.~A.~Dadashov$^{2}$}%
\affiliation{$^{1}$Department of Physics, Faculty of Arts and
Sciences, Erciyes University, Kayseri, Turkey \\
$^{2}$Baku State University, Z. Khalilov
st. 23, AZ-1148, Baku, Azerbaijan}%
\date{\today}
\begin{abstract}
In this article, we investigate the Drell-Yan process of the light
neutralino pair $\widetilde{\chi}_{i}^{0}\widetilde{\chi}_{j}^{0}$
($i, j=1, 2$) productions at proton-proton collisions and we present
the general formulae for the differential cross sections. We conduct
an extensive examination of the dependence of the total cross
section of the subprocesses $q\overline
q\rightarrow\widetilde\chi_{i}^{0}\widetilde\chi_{j}^{0}$ on the
beam energy, on the mass of the squarks and also on the $M_2$
gaugino for the three extremely different scenarios. For all three
cases, the outcomes are as follows: The dependence of the total
cross section of the subprocesses
$q\bar{q}\to\widetilde{\chi}_{i}^{0}\widetilde{\chi}_{j}^0$ on the
beam energy is dominated by one of the subprocesses,
$q\bar{q}\to\widetilde{\chi}_{1}^{0}\widetilde{\chi}_{2}^0$. On the
other hand, the dependence of the total cross section of the
subprocesses
$q\bar{q}\to\widetilde{\chi}_{i}^{0}\widetilde{\chi}_{j}^0$ on the
mass of the squarks is dominated by one of the subprocesses,
$q\bar{q}\to\widetilde{\chi}_{1}^{0}\widetilde{\chi}_{1}^0$. We
derive there from that our findings may lead to new insights
relating to experimental investigations and these dependencies may
be used as bases of an experimental research for the neutralino pair
at LHC.
\end{abstract}

\maketitle

\section{\bf Introduction}

The Standard Model (SM) is a successful theory of strong and
electroweak interactions up to the energies accessible at present
[1]. The hierarchy problem suggests that, in principle,  SM is one
of the fundamental effective theories of the low energy region.
Supersymmetry (SUSY) is presently the most popular attempt to
solve the hierarchy problem of SM, where the cancelletion of
quadratic divergences is guaranteed and hence any mass scale is
stable under radiative corrections. The most favorable candidate
for a realistic extension of SM is the minimal supersymmetric
standard model (MSSM). In MSSM, a discrete symmetry called
R-parity [2,3] is kept in order to assure baryon and lepton number
conservations since the gauge-coupling unification supports
conservation of R-parity. The minimal supersymmetric standard
model (MSSM) [4] predicts that there exists an absolutely stable
LSP. Most often the LSP in the MSSM theory is the lightest
Majorana fermionic neutralino $\widetilde{\chi}_{1}^{0}$.
Therefore the production of the lightest neutralino
$\widetilde{\chi}_{1}^{0}$ and the second lightest neutralino
$\widetilde{\chi}_{2}^{0}$  may be studied at present and future
experiments and the detailed  study of the neutralino sector will
help us to determine which kind of the supersymmetric models
really exists in nature. They are determined by diagonalizing the
corresponding mass matrix. In MSSM, the mass matrix depends on
four unknown parameters, namely $\mu$, $M_{2}$, $M_{1}$, and
$tan\beta =v_2/v_1$, which is the ratio of the vacuum expectation
values of the two Higgs fields. $\mu$ is the supersymmetric
Higgs-boson mass parameter and $M_{2}$ and $M_{1}$ are the gaugino
mass parameters associated with the $SU(2)$ and $U(1)$ subgroups,
respectively. The direct search of supersymmetric particles in
experimental research is one of the promising tasks for present
and future colliders. The multi-TeV Large Hadron Collider (LHC) at
CERN and the possible future Next Linear Collider (NLC) are
elaborately designed in order to study the symmetry-breaking
mechanism and the new physics beyond SM. If the supersymmetry
really exists at TeV scale, SUSY particles should be discovered
and it will be possible to make accurate measurements to determine
their masses and other parameters of the Lagrangian at LHC, and
then we will have a better understanding of the supersymmetry
model. We know that there are several mechanisms inducing the
production of a neutralino/chargino pair at hadron colliders. One
is through the quark-antiquark annihilation, called the Drell-Yan
process, and another is via gluon-gluon fusion. Although the
antiquark luminosity in the distribution function of the proton is
much lower than gluon, the cross sections of the neutralino pair
productions via the Drell-Yan mechanism are competitive with those
from the gluon-gluon fusion, since the former mechanism of the
neutralino pair productions is accessible at the tree level. These
facts make the production rates in the Drell-Yan process
competitive with or even larger than those in gluon-gluon fusions.
But, the reactions $q\bar{q}\to
\widetilde\chi_{i}^{0}\widetilde\chi_{j}^{0}$ are only
subprocesses of the parent $pp$ hadron collider. In work [5], the
authors have considered the production of neutralino pair at a
high energy hadron collider, putting a special emphasis on the
case where one of them is the lightest neutralino
$\widetilde{\chi}_{1}^{0}$, possibly constituting the main Dark
Matter component.  Neutralino pair production in proton-proton
collisions have been studied in [6] as well, but, our results
disagree with those conducted in Ref. [6]. In our calculations, we
have used the anticommuting nature of the Fermionic fields in
amplitude and cross section of processes, which do not agree with
the fermionic symmetry property assumed in Ref. [6]. Therefore,
this approach is significant for the theoretical and experimental
studies relating to the neutralino pair productions through the
proton-proton collisions at LHC. These results imply an
interesting complementarity between the future LHC measurements
and the related $\gamma\gamma \to
\widetilde\chi_{i}^{0}\widetilde\chi_{j}^{0}$ measurements at a
future Linear Collider. Within this context, this paper is
organized as follows: in section \ref{parameters}, we present some
formulae for the neutralino/chargino sector. In section \ref{cs},
we provide the formulae for the amplitudes and the differential
cross sections of subprocesses $q\bar{q}\to
\widetilde\chi_{i}^{0}\widetilde\chi_{j}^{0}$
 and in section \ref{results}, we present the numerical results for the cross-section
and discuss the dependence cross-section on the SUSY model
parameters. We state our conclusions in section \ref{Conc}.

\section{MSSM parameters in neutralino/chargino sector}\label{parameters}

In the minimal supersymmetric extension of the Standard Model
(MSSM), the  physical neutralino mass eigenstates $\chi_1^{0}$
(i=1,2,3,4) are the combinations of the neutral gauginos (
$\widetilde B$ and $\widetilde W^{3}$) and the neutral higgsinos
($\widetilde H_1^{0}$, $\widetilde H_2^{0}$). In the two-component
fermion fields
${\psi}_{j}^{0}=(-i\lambda^1,-i\lambda^3,{\psi}_{{H}_{1}^{0}},{\psi}_{{H}_{2}^{0}})$
[4,7],where $\lambda^1$ is the bino and $\lambda^3$ is the neutral
wino, the neutralino mass term in the Lagrangian is given by
$$
L=-\frac{1}{2}(\psi^0)^{T}M\psi^0 +h.c.
$$
 The neutralino mass matrix [4,7] in the $(\widetilde
B,\widetilde W,\widetilde H_1^{0},\widetilde H_2^{0})$ basis,
$$ M=\left(\begin{array}{cccc}M_{1}&0&-m_{Z}c_{\beta}s_{W}&m_{Z}s_{\beta}s_{W}\\
0&M_{2}&m_{Z}c_{\beta}c_{W}&-m_{Z}s_{\beta}c_{W}\\
-m_{Z}c_{\beta}s_{W}&m_{Z}c_{\beta}c_{W}&0&-\mu\\
m_{Z}s_{\beta}s_{W}&-m_{Z}s_{\beta}c_{W}&-\mu&0\end{array}\right)
$$
is built up by the fundamental supersymmetry parameters: the $U(1)$
and $SU(2)$ gaugino masses $M_{1}$ and $M_{2}$, the higgsino mass
parameter $\mu$, and the ratio $tan\beta =v_2/v_1$ of the vacuum
expectation values of the two neutral Higgs fields, which break the
electroweak symmetry. Here, $s_{\beta}= sin\beta$,
$c_{\beta}=cos\beta$ and $s_{W}$, $c_{W}$ are the sine and cosine of
the electroweak mixing angle $\theta_{W}$. In CP-noninvariant
theories, the mass parameters are complex. By the reparametrization
of the fields, $M_{2}$ can be taken as real and positive without
loss of generality so that the two remaining nontrivial phases,
which are reparametrization-invariant, may be attributed to $M_{1}$
and $\mu$:
$$
M_{1} =|M_{1}|e^{i\phi_{1}}\,\,\,\,\, and \,\,\,\,
\mu=|\mu|e^{i\phi_{\mu}}\,\,\, (0\leq\phi_1,\phi_\mu<2\pi)
$$
The experimental analysis of neutralino properties in production and
decay mechanisms will unravel the basic structure of the underlying
supersymmetries theory. The charginos $\widetilde\chi_j^{+}$ ($j=1,
2$) mass matrix in the current eigenstate basis have the form [7]
$$
M_c=\left(\begin{array}{cc}M_2&{\sqrt{2}}m_Wc_\beta \\
\sqrt{2}m_Ws_\beta &|\mu|e^{i\phi_\mu}\end{array}\right)
$$
is diagonalized by two different unitary matrices $U_R M_c U_L^{+}
=diag\left\{{{m_1^{\pm},m_2^{\pm}}}\right\}$, parametrized in
general by two rotation angles and four phases:
$$
U_L=\left(\begin{array}{cc}c_L&s_L^{\star}\\
-s_L&c_L\end{array}\right)
$$
and
$$
U_R= diag\{e^{i\gamma_1},e^{i\gamma_2}\}\cdot\left(\begin{array}{cc}c_R&s_R^{\star}\\
-s_R&c_R\end{array}\right)
$$
 where
$c_{L,R} =cos\phi_{L,R}$ \,\,\, and
$s_{L,R}=sin\phi_{L,R}e^{i\delta_{L,R}}$ . In the limit of
$M_{2}^2$,$|\mu|^2\gg m_{Z}^2$\,\,\, and $|M_2\pm|\mu||^2 \gg
m_{Z}^2$, the following expressions
$$
m_{1}^{\pm}=M_2+X_2[M_2+|\mu|s_{2\beta}cos\phi_\mu],
$$
$$
m_{2}^{\pm}=|\mu|X_2[|\mu|+M_2s_{2\beta}cos\phi_\mu],
$$
are found for the chargino masses and
$$
s_L=\frac{\sqrt{2}m_W}{M_2^2-|\mu|^2}(Mc_\beta + \mu^\star
s_{\beta}) \,\,\,\,\,\,\,\,
\gamma_1=+X_2\frac{|\mu|}{M_2}s_{2\beta}sin\phi_\mu$$
$$
s_R=\frac{\sqrt{2}m_W}{M_2^2-|\mu|^2}(\mu c_\beta + M_{2}^\star
s_{\beta}) \,\,\,\,\,\,\,\,
\gamma_2=-X_2\frac{M_2}{|\mu|}s_{2\beta}sin\phi_\mu
$$
$$
X_2=\frac{m_{Z}^2 c_{W}^2}{|M_2|^2-|\mu|^2}
$$
 for the mixing angles and phases. In the present work, we have
investigated the Drell-Yan process of the light neutralino pair
$\widetilde{\chi}_{i}^{0}\widetilde{\chi}_{j}^{0}$ ($i, j=1, 2$)
productions at hadron colliders. Since the neutralino mass matrix
$M$ is symmetric, one unitary matrix $N$ is sufficient to rotate the
gauge eigenstate basis ($\widetilde{B}^0,\widetilde{W}^3,\widetilde
{H}_{1}^0,\widetilde {H}_{2}^0$) to the mass eigenstate basis of the
Majorana fields $\widetilde{\chi}_{i}^{0}$.

\begin{equation}
M_D=N^{T}MN=\sum_{j=1}^{4}m_{\widetilde{\chi}_{j}^{0}}E_{j},
\end{equation}

where $N$ is a unitary matrix. To determine $N$, it is easiest to
square eq.(2.1) obtaining
\begin{equation}
M_{D}^2=N^{-1}M^{+}MN=\sum_{j=1}^{4}m_{\widetilde{\chi}_{j}^{0}}^2E_{j},
\end{equation}
where $(E_j)_{4x4}$ are the basic matrices defined by
$(E_j)_{ik}=\delta_{ji}\delta_{jk}$ and $\widetilde{\chi}_{j}^{0}$
stand for the four component Majorana neutralinos:

\begin{equation}
\widetilde{\chi}_{j}^{0}=\left(\begin{array}{ccc}\widetilde{\chi}_{j}^{0}\\...\\
\widetilde{\chi}_{j}^{0}\\ \end{array}\right), j=1,....,4
\end{equation}
Here, we suppose that the real eigenvalues of $M_D$ are ordered in
the following way
$$
m_{\widetilde{\chi}_{1}^{0}}\leq m_{\widetilde{\chi}_{2}^{0}}\leq
m_{\widetilde{\chi}_{3}^{0}}\leq m_{\widetilde{\chi}_{4}^{0}}.
$$
The mass eigenvalues $m_{\widetilde{\chi}_{j}^{0}}$ ($j=1,2,3,4$) in
$M_D$ can be chosen as positive by a suitable definition of the
unitary matrix $N$. In this work, we consider the higgsino/gaugino
sector with the following assumptions: First, for simplification,
$CP$-conservation is hold, namely $\phi_{\mu}=\phi_1=0$. The
physical signs among $M_1$, $M_2$ and $\mu$ are relative, which can
be absorbed into phases $\phi_{\mu}$ and $\phi_1$ by redefinition of
fields. Thus, $M_1$, $M_2$ and $\mu$ are chosen to be real and
positive, \emph{i.e.}, $M_1$, $M_2$, $\mu>0$. With the above
assumptions, there are several scenario for the choice of the SUSY
parameters for the investigation of the neutralino pair production
in hadron collider. One can employ the scenario of taking $M_1$,
$M_2$, $\mu$, and $\tan\beta$ as input parameters, and then get all
the physical chargino and neutralino masses and the matrix elements
of $U_R$, $U_{L}^{+}$ and $N$ as outputs. Also, there are other
alternative scenarios, such as the CP conserving mSUGRA scenario
with five input parameters, namely $m_{1/2}$, $m_{0}$, $A_{0}$,
$\mu$ and $\tan\beta$, where $m_{1/2}$, $m_{0}$ and $A_{0}$ are the
universal gaugino mass, scalar mass at GUT scale and the trilinear
soft breaking parameter in the superpotential respectively. From
these five parameters, all the masses and couplings of the model are
determined by the evolution from the GUT scale down to the low
electroweak scale [8]. Since SUSY parameters should be extracted
from the physical quantities, one can also choose an alternative way
to diagonalize the mass matrix $M$, by taking any two physical
chargino masses together with $\tan\beta$ as inputs. There are
several scenarios about the choice of two chargino masses and
$tan\beta$ [10]. Also there are two possible scenario about the
choice of $tan\beta$: scenario with small
$tan\beta$($tan\beta\approx1\div3$) and scenario with large
$tan\beta$($tan\beta\approx30\div70$)[9]. In this work, we take two
chargino masses $m_{\widetilde{\chi}_{1,2}^{+}}$ and scenario with
small $tan\beta$ as inputs. In this way, the two fundamental SUSY
parameters, $M_2$ and $\mu$ can be figured out from the chargino
masses by using the following formula: For given $tan\beta$, the
fundamental SUSY parameters $M_2$ and $\mu$ can be derived from
these two chargino masses [11]. The sum and differences of the
chargino masses lead to the following equations involving $M_2$ and
$\mu$:

\begin{equation}
M_{2}^2+{|\mu|}^2=m_{\widetilde{\chi}_{1}^{+}}+m_{\widetilde{\chi}_{2}^{+}}-2m_{W}^2,
\end{equation}

\begin{equation}
M_{2}^2|\mu|^2-2m_{W}^2\sin2\beta cos\phi_{\mu}M_2|\mu|+(m_{W}^4
sin^2{2\beta}-m_{\widetilde{\chi}_{1}^{+}}^2m_{\widetilde{\chi}_{2}^{+}}^2)=0.
\end{equation}

The solution of (2.5) is given as:

\begin{equation}
M_2|\mu|=m_{W}^2cos\phi_{\mu}sin2\beta \pm
\sqrt{m_{\widetilde{\chi}_{1}^{+}}^2
 m_{\widetilde{\chi}_{2}^{+}}^2
-m_{W}^4sin^2{2\beta} sin^2{\phi_{\mu}}}.
\end{equation}

From (2.4) and (2.6), one obtains the following solutions for $M_2$
and $\mu$:

\begin{equation}
2M_2^2=(m_{\widetilde{\chi}_{1}^{+}}^2+m_{\widetilde{\chi}_{2}^{+}}^2-
2m_{W}^2)\mp
\left(\sqrt{(m_{\widetilde{\chi}_{1}^{+}}^2+m_{\widetilde{\chi}_{2}^{+}}^2-
2m_{W}^2)^2-\Delta_{\pm} }\right),
\end{equation}

\begin{equation}
2|\mu|^2=(m_{\widetilde{\chi}_{1}^{+}}^2+m_{\widetilde{\chi}_{2}^{+}}^2-
2m_{W}^2)\pm
\left(\sqrt{(m_{\widetilde{\chi}_{1}^{+}}^2+m_{\widetilde{\chi}_{2}^{+}}^2-2m_{W}^2)^2-
\Delta_{\pm}}\right ),
\end{equation}
with
$$
\Delta_{\pm}=4 \left[m_{\widetilde{\chi}_{1}^{+}}^2
m_{\widetilde{\chi}_{2}^{+}}^2+ m_{W}^4
cos2\phi_{\mu}sin^2{2{\beta}}\pm 2m_{W}^2 cos\phi_{\mu}
sin2\beta\cdot \right.
$$
$$
\left.\sqrt{m_{\widetilde{\chi}_{1}^{+}}^2
m_{\widetilde{\chi}_{2}^{+}}^2-m_{W}^4 sin^2{2{\beta}}
sin^2{\phi_{\mu}}}\right],
$$
where the upper signs correspond to $M_2<|\mu|$ regime, and the
lower ones to $M_2>|\mu|$. Therefore, for given $tan\beta$, $M_2$
and $\mu$ can be determined in terms of the masses of the charginos
$m_{\widetilde{\chi}_{1}^{+}}$ and $m_{\widetilde{\chi}_{2}^{+}}$ by
using (2.7), and (2.8) from which one gets four solutions
corresponding to different physical scenarios. For $\mu<M_{2}$, the
lightest chargino has a stronger higgsino-like component and
therefore is referred to as higgsino-like. The solution $\mu>M_{2}$,
corresponding to the gaugino-like situation, can be readily obtained
by the substitutions: $M_{2} \to \mu$, and $\mu \to$
sign($\mu$)$M_{2}$. In this paper, we assume the GUT relation
[11,12]
\begin{equation}
M_1=\frac{5}{3}M_2 tan^2\Theta_W.
\end{equation}
Thus, the neutralino masses can be determined by solving the
characteristic equation associated to this system, that is

\begin{equation}
X^4-aX^3+bX^2-cX+d=0,
\end{equation}
where
$$
a =  M_1^2+2\mu^2+M_2^2+2m_Z^2,
$$
$$
b =
(\mu^2+m_Z^2)^2+M_2^2(M_1^2+2\mu^2+2m_{Z}^2s_W^2)+2M_1^2(\mu^2+m_Z^2c_W^2)-2\mu
m_Z^2c_W^2 M_2sin2\beta
$$
$$
\times cos\phi_{\mu}-2m_Z^2s_W^2M_1 sin2\beta
cos(\phi_{\mu}+\phi_1),
$$
$$
c =  {\mu}^4 M_{1}^2+{\mu}^2 m_{Z}^4 sin^2
2\beta+M_1^2m_Z^2c_W^2(2\mu^2+m_Z^2c_W^2)+
$$
$$
M_2^2(m_{Z}^4s_{W}^4+2\mu^2(m_Z^2s_W^2+M_1^2)+\mu^4)-2\mu
m_Z^2s_W^2M_1(\mu^2+M_2^2)sin2\beta cos(\phi_{\mu}+\phi_1)+
$$
$$
2m_Z^2 c_W^2 M_2[m_Z^2M_1s_W^2cos\phi_1-\mu(\mu^2+M_1^2)
cos\phi_{\mu} sin2\beta],
$$
$$
d =  m_Z^4 c_W^4\mu^2 M_1^2
sin^2{2\beta}+2m_Z^2\mu^2M_1M_2c_W^2(m_Z^2s_W^2sin2\beta
cos\phi_1-\mu M_1cos\phi_\mu) +
$$
$$
\mu^2m_Z^2s_W^2M_2^2sin2\beta(m_Z^2s_W^2sin2\beta-2\mu
M_1cos(\phi_1+\phi_{\mu}))+\mu^4 M_1^2M_2^2.
$$
Solving Eq.(2.10), we get the exact analytic formulae for the
neutralino masses
$$
m_{\widetilde{\chi}_{1}^{0}}^2,
m_{\widetilde{\chi}_{2}^{0}}^2=\frac{a}{4}-\frac{f}{2}\mp\frac{1}{2}\sqrt{r-w-\frac{p}{4f}},
$$
\begin{equation}
m_{\widetilde{\chi}_{3}^{0}}^2,
m_{\widetilde{\chi}_{4}^{0}}^2=\frac{a}{4}+\frac{f}{2}\mp\frac{1}{2}\sqrt{r-w+\frac{p}{4f}},
\end{equation}
where
$$
f=\sqrt{\frac {r}{2}+w},
$$
$$
w=\frac{q}{(3\cdot 2^{1/3})}+\frac{(2^{1/3}\cdot h)}{3\cdot q},
$$
$$
q=(k+{\sqrt{k^2-4h^3}})^{1/3},
$$
\begin{equation}
k=2b^3-9abc+27c^2+27a^2d-72bd,
\end{equation}
$$
h=b^2-3ac+12d,
$$
$$
p=a^3-4ab+8c,
$$
$$
r=\frac{a^2}{2}-\frac{4b}{3}.
$$
Starting from Eq.(2.2), we get
\begin{equation}
(M^{+}M)N-NM_{D}^2=0.
\end{equation}
A more explicit form of this matrix equation is
$$
(A_{11}-m_{\widetilde{\chi}_{j}^{0}}^2)N_{1j}+A_{12}N_{2j}+A_{13}N_{3j}+A_{14}N_{4j}=0,
$$
$$
A_{21}N_{1j}+(A_{22}-m_{\widetilde{\chi}_{j}^{0}}^2)N_{2j}+A_{23}N_{3j}+A_{24}N_{4j}=0,
$$
$$
A_{31}N_{1j}+A_{32}N_{2j}+(A_{33}-m_{\widetilde{\chi}_{j}^{0}}^2)N_{3j}+A_{34}N_{4j}=0,
$$
\begin{equation}
A_{41}N_{1j}+A_{42}N_{2j}+A_{43}N_{3j}+(A_{44}-m_{\widetilde{\chi}_{j}^{0}}^2)N_{4j}=0,
\end{equation}
 $j$=1,...,4, where $A_{ij}=\sum_{k=1}^{4}M_{ki}^{\star}M_{kj}$:
\begin{eqnarray}
A_{11} &= & M_{1}^2+m_{Z}^2 s_{W}^2 \nonumber,\\
A_{12}&=&A_{21}=-m_{Z}^2 s_{W}c_{W} \nonumber,\\
A_{13}&=&A_{31}=-M_1m_Zc_{\beta}s_W-\mu m_Zs_Ws_{\beta}\nonumber,\\
A_{14}&= &A_{41}=M_1m_Zs_{\beta}s_W+\mu m_Zc_{\beta}s_W \nonumber,\\
A_{22}&= &M_2^2+m_Z^2c_W^2 \nonumber,\\
A_{23}&=&A_{32}=M_2m_Zc_{\beta}c_W+\mu m_Zs_{\beta}c_W \nonumber,\\
A_{33}&= &\mu^2+m_Z^2c_{\beta}^2 \nonumber,\\
A_{24}&= &A_{42}=-M_2m_Zs_{\beta}c_W-\mu m_Zc_{\beta}c_W \nonumber,\\
A_{34}&= &A_{43}=-m_Z^2s_{\beta}c_{\beta} \nonumber, \\
A_{44}&= &m_Z^2 s_{\beta}^2+\mu^2 \nonumber.
\end{eqnarray}
The diagonalizing matrix $N$ can be obtained by computing the
eigenvectors corresponding to the eigenvalues given in Eq.(2.11).
Indeed, by inserting a generic eigenvalue
$m_{\widetilde{\chi}_{j}^{0}}$, into Eq.(2.14) and dividing each one
of these equations by $N_{1j}$, where it is assumed that
$N_{1j}\not=0$, we get
$$
A_{12}\frac{N_{2j}}{N_{1j}}+A_{13}\frac{N_{3j}}{N_{1j}}+A_{14}\frac{N_{4j}}{N_{1j}}-m_{\widetilde{\chi}_{j}^{0}}^2=-A_{11},
$$
$$
(A_{22}-m_{\widetilde{\chi}_{j}^{0}}^2)\frac{N_{2j}}{N_{1j}}+A_{23}\frac{N_{3j}}{N_{1j}}+A_{24}\frac{N_{4j}}{N_{1j}}=-A_{21},
$$
$$
A_{32}\frac{N_{2j}}{N_{1j}}+(A_{33}-m_{\widetilde{\chi}_{j}^{0}}^2)\frac{N_{3j}}{N_{1j}}+A_{34}\frac{N_{4j}}{N_{1j}}=-A_{31},
$$
\begin{equation}
A_{42}\frac{N_{2j}}{N_{1j}}+A_{43}\frac{N_{3j}}{N_{1j}}+(A_{44}-m_{\widetilde{\chi}_{j}^{0}}^2)\frac{N_{4j}}{N_{1j}}=-A_{41},
\end{equation}
Solving this system of equations, and taking into account the
relation
\begin{equation}
|N_{1j}|^2+|N_{2j}|^2+|N_{3j}|^2+|N_{4j}|^2=1,
\end{equation}
it yields the $N_{ij}$ matrix's component
\begin{equation}
N_{ij}=\frac{{\Delta}_{ij}}{{\Delta}_{1j}}\cdot
\frac{|{{\Delta}_{1j}}|}{\sqrt{|{\Delta}_{1j}|^2+|{\Delta}_{2j}|^2+|{\Delta}_{3j}|^2+|{\Delta}_{4j}|^2}},
\end{equation}
when, $i=1,...4$. Here,
$$
\Delta_{1j}=\left|\begin{array}{ccc}A_{22}-m_{\widetilde{\chi}_{j}^{0}}^2&A_{23}&A_{24}\\
A_{32}&A_{33}-m_{\widetilde{\chi}_{j}^{0}}^2&A_{34}\\
A_{42}&A_{43}&A_{44}-m_{\widetilde{\chi}_{j}^{0}}^2\end{array}\right|
$$
and $\Delta_{ij},i=2,3,4,$ is formed from $\Delta_{1j}$ by
substituting the (i-1)th column by  $\left(\begin{array}{ccc}\\-A_{21}\\-A_{31}\\
-A_{41}\\ \end{array}\right)$.
\section{Calculation of the cross section }\label{cs}
The Neutralino pair productions, which can be produced via the
collisions of quark and antiquarks in protons, can be expressed as
\begin{equation}
q(p_1)\overline
q(p_2)\rightarrow\widetilde\chi_{i}^{0}(k_1)\widetilde\chi_{j}^{0}(k_2),
\end{equation}
where $p_{1}$ and $p_{2}$ represent the momenta of the incoming
quark and antiquark, and $k_1$ and $k_2$ denote the momenta of the
two final state neutralinos, respectively. The Mandelstam invariant
variables for subprocess (3.1) are defined as
\begin{equation}
\hat s=(p_1+p_2)^2, \quad \hat t=(p_1-k_1)^2,\quad \hat
u=(p_1-k_2)^2.
\end{equation}
The Feynman diagrams of the subprocess are shown in Fig.1. The
relevant couplings of the supersymmetric particles are deduced from
the following interaction Lagrangians of the Supersymmetric Standard
Model [13]:
\begin {equation}
L_{Z^{0}\widetilde{\chi}_{i}^{0}\widetilde{\chi}_{j}^{0}}=
\frac{1}{2}\frac{g}{cos\Theta_{\rm W}}
Z_{\mu}\overline{\widetilde{\chi}}_{i}^{0}\gamma^{\mu}(O_{ij}^{\prime\prime}
L_{q} P_{L}+O_{ij}^{\prime\prime} R_{q}
P_{R})\widetilde{\chi}_{j}^{0},
\end {equation}
\begin {equation}
L_{Z^{0}q\bar{q}}=\frac{g}{cos\Theta_{\rm W}} \bar {q}\gamma^{\mu}(
L_{q} P_{L}+R_{q} P_{R})q Z_{\mu},
\end {equation}
\begin {equation}
L_{q\widetilde{q}\widetilde{\chi}^{0}}=-\sqrt
{2}g\bar{q}[a_{i}^{L}(\widetilde{q}_{n})P_L
+a_{i}^{R}(\widetilde{q}_{n})P_R]\widetilde{\chi}_{i}^{0}\widetilde{q}_{n}.
\end {equation}
In Eqs.(3.3-3.5) $\widetilde{\chi}_{i}^{0}$ and $q$ are
four-component spinor fields and $\widetilde{q}$ is the field of the
squark. Furthermore, $g=e/sin\Theta_{W}$ $(e>0)$ is the weak
coupling constant, $\Theta_W$ the Weinberg angle,
$P_{R,L}=\frac{1}{2}(1\pm\gamma^5)$, while the coupling constant
$O_{ij}^{\prime\prime}, L_{q},R_{q}$ and
$a_{i}^{R,L}(\widetilde{q}_{n})$ are given by
\begin{equation}
O_{ij}^{{\prime\prime}
L}=\frac{1}{2}(N_{i3}N_{j3}^{\star}-N_{i4}N_{j4}^{\star})cos2\beta
-\frac{1}{2}(N_{i3}N_{j4}^{\star}+N_{i4}N_{j3}^{\star})sin2\beta,
\end{equation}
\begin{equation}
O_{ij}^{{\prime\prime} R}=-O_{ij}^{{\prime\prime} L\star},
\end{equation}
\begin{equation}
L_{q}=2I_{q}^3(1-2sin^2\Theta_{W}|Q_q|),\,\,\,R_{q}=-2sin^2\Theta_{W}Q_q,
\end{equation}
with $I_{q}^3,Q_q$ being the isospin and charge of the various
$q_{L}$-quarks, and
$$
a_i^L(\widetilde{u}_L)=-\frac{e}{{3\sqrt 2}s_W
c_W}(N_{1i}s_W+3N_{2i}c_W),
$$
$$
a_i^R(\widetilde{u}_R)=\frac{2{\sqrt 2}e}{3 c_W}N_{1i}^{\star},
$$
$$
a_i^L(\widetilde{d}_L)=-\frac{e}{{3\sqrt 2}s_W c_W}(N_{1i}
s_W-3N_{2i} c_W),
$$
$$
a_i^R(\widetilde{d}_R)=-\frac{{\sqrt 2}e}{3 c_W} N_{1i}^{\star},
$$
$$
a_i^L(\widetilde{u}_R)=-\frac{e m_u}{{\sqrt 2} m_W s_W
s_{\beta}}N_{4i},
$$
$$
a_i^R(\widetilde{u}_R)=-\frac{e m_u}{{\sqrt 2} m_W s_W
s_{\beta}}N_{4i}^{\star},
$$
$$
a_i^L(\widetilde{d}_R)=-\frac{e m_d}{{\sqrt 2} m_W s_W
c_{\beta}}N_{3i},
$$
\begin{equation} a_i^R(\widetilde{d}_L)=-\frac{e m_d}{{\sqrt 2} m_W
s_W c_{\beta}}N_{3i}^{\star}.
\end{equation}
In (3.9), $(q=u,d)$ refer to the incoming up and down quark
(antiquark) of any family, while
$(\widetilde{q}_n=\widetilde{q}_L,\widetilde{q}_R)$ denote the
corresponding squarks. We also note that the mixing matrices N in
(3.6, 3.9), control the Bino, Wino, Higgsino components of the
neutralino in the
$Z\widetilde{\chi}_{i}^{0}\widetilde{\chi}_{j}^{0}$ and
$q\widetilde{q}\widetilde{\chi}^{0}$ coupling. The corresponding
Lorentz invariant matrix element for each of the diagrams can be
written as
\begin{equation}
T=T_{\hat s} +T_{\hat t} +T_{\hat u},
\end{equation}
where
$$
 T_{\hat s}=-\frac{e^2}{2sin^2\Theta_W cos^2\Theta_W}D_{Z}(\hat s)\overline{u}_{i}(k_1)
\gamma{\mu}[O_{Z}^{ij}P_L - O_{Z}^{ij\star}P_R]{\vartheta}_{j}
(k_2)\times
$$
$$
\overline{v}(p_2)\gamma_{\mu}(g_{V_{q}}+g_{A_{q}}\gamma_5)u(p_1),
$$
$$
T_{\hat t}=\sum_{n}\frac{1}{\hat t-m_{\widetilde
q_n}^2}\bar{u}_{i}(k_1)(a_{i}^{L}(\widetilde{q}_{n})P_L+a_{i}^{R}(\widetilde{q}_{n})P_R)u(p_1)\bar{v}(p_2)\times
$$
\begin{equation}
(a_{j}^{L\star}(\widetilde{q}_{n})P_L
+a_{j}^{R\star}(\widetilde{q}_{n})P_R)v_{j}(k_2),
\end{equation}
$$
T_{\hat u}=-\sum_{n}\frac{1}{\hat u - m_{\widetilde q_n}^2}
\bar{u}_{j}(k_2)(a_{j}^{L\star}(\widetilde{q}_{n})P_R +
a_{j}^{R\star}(\widetilde{q}_{n})P_L)u(p_1)\bar{v}(p_2)\times
$$
$$
(a_{i}^{L}(\widetilde{q}_{n})P_L + a_{i}^{R}(\widetilde{q}_{n})P_R)
v_{i}(k_1),
$$
where the index $n$ refers to the summation over the exchanged $L$-
and $R$- squarks of the same flavor in the $t$-and $u$- channel, and
$(i,j)$ describe the final neutralinos. From the total
(gauge-invariant) amplitude $T$, which is the sum of the partial
amplitudes Eq.(3.11), we obtain the differential cross section as

\begin{equation}
\frac{d\sigma}{d\Omega}=\frac{\lambda_{ij}}{384\pi^2\hat
s^2}(\frac{1}{2})^{\delta_{ij}}(M_{\hat s \hat s}+M_{\hat t \hat
t}+M_{\hat u \hat u} -2M_{\hat s \hat t} +2M_{\hat s \hat
u}-2M_{\hat t \hat u}),
\end{equation}
where
\begin{equation}
\lambda_{ij}=\sqrt{(\hat s-m_{\widetilde{\chi}_{i}^{0}}^2-
m_{\widetilde{\chi}_{j}^{0}}^2)^2-4m_{\widetilde{\chi}_{i}^{0}}^2m_{\widetilde{\chi}_{j}^{0}}^2}/2,
\end{equation}
and $(\frac{1}{2})^{\delta_{ij}}$ is the final identical-particle
factor. The squares of the matrix element have the form as
$$
M_{\hat s \hat s}=\frac{e^4}{4\sin^4\Theta_W
\cos^4\Theta_{W}}|D_{Z}(\hat s)|^2 (L_{q}^2+R_{q}^2)
{O_{Z}^{ij}O_{Z}^{ij \star}}[(m_{\widetilde{\chi}_{i}^{0}}^2 - \hat
u)(m_{\widetilde{\chi}_{j}^{0}}^2 -\hat u)+
$$
\begin{equation}
(m_{\widetilde{\chi}_{i}^{0}}^2 - \hat
t)(m_{\widetilde{\chi}_{j}^{0}}^2 -\hat
t)-m_{\widetilde{\chi}_{i}^{0}}m_{\widetilde{\chi}_{j}^{0}}\hat s
(O_{Z}^{ij2}+O_{Z}^{ij \star 2})],
\end{equation}
$$
M_{\hat t \hat t}=\frac{1}{(\hat t - m_{\widetilde {q}_{k}}^2)(\hat
t-m_{\widetilde{q}_{l}}^2)}(a_{i}^{L}(\widetilde{q}_{k})a_{i}^{L\star}
(\widetilde{q}_{l})+a_{i}^{R}(\widetilde{q}_{k})a_{i}^{R\star}(\widetilde{q}_{l}))
(a_{j}^L(\widetilde{q}_{k})a_{j}^{L\star}(\widetilde{q}_{l}) + \\
$$

\begin{equation}
a_{j}^R(\widetilde{q}_{k})a_{j}^{R\star}(\widetilde{q}_{l}))
(m_{\widetilde{\chi}_{i}^{0}}^2- \hat
t)(m_{\widetilde{\chi}_{j}^{0}}^2 - \hat t),
\end{equation}

$$
M_{\hat u \hat u}=\frac{1}{(\hat u - m_{\widetilde{q}_{k}}^2)(\hat
u-m_{\widetilde{q}_{l}}^2)}(a_{i}^{L\star}(\widetilde{q}_{k})a_{i}^L
(\widetilde{q}_{l})+a_{i}^{R\star}(\widetilde{q}_{k})a_{i}^R(\widetilde{q}_{l}))
(a_{j}^L(\widetilde{q}_{l}) a_{j}^{L\star}(\widetilde{q}_{k}) + \\
$$

\begin{equation}
a_{j}^R(\widetilde{q}_{l}) a_{j}^{R\star}(\widetilde{q}_{k}))
(m_{\widetilde{\chi}_{i}^{0}}^2- \hat
u)(m_{\widetilde{\chi}_{j}^{0}}^2 - \hat u),
\end{equation}

$$
M_{\hat t \hat u}=\frac{1}{(\hat t -
{m}_{\widetilde{q}_{k}}^2)(\hat
u-{m}_{\widetilde{q_l}}^2)}\biggl\{\frac{1}{2}\left
[a_{i}^{L\star}(\widetilde{q}_{k})a_{j}^L(\widetilde{q}_{l})
a_{j}^R(\widetilde{q}_{k})
a_{i}^{R\star}(\widetilde{q}_{l})+a_{i}^{R\star}(\widetilde{q}_{k})
a_{j}^R(\widetilde{q}_{l}) \right.
$$
$$
\left.a_{i}^{L\star}(\widetilde{q}_{l})
a_{j}^L(\widetilde{q}_{k})\right]
((m_{\widetilde\chi_{j}^{0}}^2-\hat
u)(m_{\widetilde{\chi}_{i}^{0}}^2-\hat u)+
(m_{\widetilde{\chi}_{j}^{0}}^2-\hat
t)(m_{\widetilde{\chi}_{i}^{0}}^2-\hat t)-\hat s(\hat s
-m_{\widetilde{\chi}_{i}^{0}}^2-m_{\widetilde{\chi}_{j}^{0}}^2)) +
m_{\widetilde{\chi}_{i}^{0}}^2 m_{\widetilde{\chi}_{j}^{0}}^2
$$
\begin{equation}
\times \hat s[a_{j}^{L\star}(\widetilde{q}_{l})
a_{i}^{L}(\widetilde{q}_{k}) a_{i}^{L}(\widetilde{q}_{l})
a_{j}^{L\star}(\widetilde{q}_{k})+
a_{j}^{R\star}(\widetilde{q}_{l}) a_{i}^{R}(\widetilde{q}_{k})
a_{i}^{R}(\widetilde{q}_{l})
a_{j}^{R\star}(\widetilde{q}_{k})]\biggr\}
\end{equation}
$$
M_{\hat s \hat u}=\frac{e^2}{2sin^2\Theta_Wcos^2\Theta_W(\hat
u-m_{\widetilde {q}_{k}}^2)}\biggl\{(Re[D_{Z}(\hat s)])[L_q
a_{i}^{L\star}(\widetilde{q}_{k})a_{j}^{L}(\widetilde{q}_{k})
O_{Z}^{ij\star}- \\
$$
$$
R_q a_{i}^{R\star} (\widetilde{q}_{k}) a_{j}^{R}(\widetilde{q}_{k})
O_{Z}^{ij}] (m_{\widetilde{\chi}_{i}^{0}}^2-\hat
u)(m_{\widetilde{\chi}_{j}^{0}}^2 -\hat u)+ [R_q
a_{i}^{R\star}(\widetilde{q}_{k}) a_{j}^{R}(\widetilde{q}_{k})
O_{Z}^{ij\star}- \\
$$
\begin{equation}
L_q a_{i}^{L\star}(\widetilde{q}_{k}) a_{j}^{L}(\widetilde{q}_{k})
O_{Z}^{ij}] m_{\widetilde{\chi}_{i}^{0}}
m_{\widetilde{\chi}_{j}^{0}} \hat s \biggr\},
\end{equation}
$$
M_{\hat s \hat t}=\frac{e^2}{2sin^2\Theta_Wcos^2\Theta_W(\hat
t-m_{\widetilde q_k}^2)}\biggl\{(Re[D_{Z}(\hat s)])[R_q
a_{j}^{R\star}(\widetilde{q}_{k})a_{i}^{R}(\widetilde{q}_{k})
O_{Z}^{ij\star}- \\
$$
$$
L_q a_{j}^{L\star}(\widetilde{q}_{k}) a_{i}^{L}(\widetilde{q}_{k})
O_{Z}^{ij}] (m_{\widetilde{\chi}_{i}^{0}}^2-\hat
t)(m_{\widetilde{\chi}_{j}^{0}}^2-\hat t)+ [L_q
a_{j}^{L\star}(\widetilde{q}_{k})
a_{i}^{L}(\widetilde{q}_{k}) O_{Z}^{ij\star}- \\
$$
\begin{equation}
R_q a_{j}^{R\star}(\widetilde{q}_{k}) a_{i}^{R}(\widetilde{q}_{k})
O_{Z}^{ij}]m_{\widetilde{\chi}_{i}^{0}}
m_{\widetilde{\chi}_{j}^{0}} \hat{s}\biggr\},
\end{equation}
The following abbreviation has been used
$$D_{Z}(\hat s^2)=\frac{1}{\hat s^2 - m_{Z}^2+im_{Z} \Gamma_{Z}}.$$
For calculation, we assume $m_{Z}$ = 91.1887 GeV and the widths of
the gauge boson by $\Gamma_{Z}=2.499947 GeV$. The basic parton model
expression for the hadron-hadron collision $h_1(p_1)h_2(p_2)\to
\widetilde{\chi}_{i}^{0}(k_i)\widetilde{\chi}_{j}^{0}(k_j)$, [14,15]
is
$$
d\sigma({h_{1}(p_1)h_{2}(p_2) \to
\widetilde{\chi}_{i}^{0}}(k_i)\widetilde{\chi}_{j}^{0}(k_j))=
\sum\int\int dx_{1} dx_{2} G_{{q_{1}}/{h_{1}}}(x_{1},Q)
 G_{{q_{2}}/{h_{2}}}(x_{2},Q)\cdot
$$
\begin{equation}
d\sigma(q_1 q_2 \to
\widetilde{\chi}_{i}^{0}\widetilde{\chi}_{j}^{0})\frac{1}{1+\delta_{q_1
q_2}}
\end{equation}
with $\widetilde{\chi}_{i}^{0}$, $\widetilde{\chi}_{J}^{0}$ being
the two produced massive particles of mass
$m_{\widetilde{\chi}_{i}^{0}}$, $m_{\widetilde{\chi}_{j}^{0}}$. Here
$G_{q_1/h_{1}}(x_1,Q)$ is the distribution function of partons of
type $(q_1=q,\bar{q})$, in the hadron of type $h_1$ at a
factorization scale Q. Taking the $h_{1}h_{2}$-c.m. system as the
lab-system, the lab-momenta of the produced
$\widetilde{\chi}_{i}^{0}$ and $\widetilde{\chi}_{j}^{0}$ are [16]
\begin{equation}
k_{i}^{\mu}=(E_i,k_{T}, k_{i}cos\theta),\,\,\,
k_{j}^{\mu}=(E_j,-k_T, k_{j} cos\theta),
\end{equation}
where their transverse momenta are obviously just opposite
\begin{equation}
                  k_T=k_{T_i}=-k_{T_j},
\end{equation}
while their transverse energies
$E_{T_i}=\sqrt{k_{T}^2+m_{\widetilde{\chi}_{i}^{0}}^2}$,
$E_{T_j}=\sqrt{k_T^2+m_{\widetilde{\chi}_{j}^{0}}^2}$ are used to
define

\begin{equation}
 x_{T_i}=\frac{2E_{T_i}}{\sqrt s}\,\,\,,\beta_{T_i}={k_T}/{E_{T_i}}=
 \sqrt{1-\frac{4m_{\widetilde{\chi}_{i}^{0}}^2}{sx_{T_i}^2}},
\end{equation}
\begin{equation}
x_{T_j}=\frac{2E_{T_j}}{\sqrt
s}\,\,\,,\beta_{T_j}={k_T}/{E_{T_j}}=\sqrt{1-\frac{4m_{\widetilde{\chi}_{j}^{0}}^2}{sx_{T_j}^2}},
\end{equation}
Note that
\begin{equation}
E_{T_j}^2=E_{T_i}^2+m_{\widetilde{\chi}_{j}^{0}}^2-m_{\widetilde{\chi}_{i}^{0}}^2,\,\,\,
x_{T_j}^2=x_{T_i}^2+4\cdot\frac{(m_{\widetilde{\chi}_{j}^{0}}^2-m_{\widetilde{\chi}_{i}^{0}}^2)}{s}
\end{equation}
The rapidites and production angles of $\widetilde{\chi}_{i}^{0}$,
$\widetilde{\chi}_{j}^{0}$, in the lab-system, are related to their
energies and momenta along the beam-axis of hadron $h_1$, by
\begin{equation}
y_i=\frac{1}{2}ln\frac{E_{i}+k_{i} cos\theta_i}{E_{i}-k_{i}
cos\theta_i},\,\,\, y_j=\frac{1}{2}ln\frac{E_{j}+k_{j}
cos\theta_j}{E_{j} -k_{j} cos\theta_j}
\end{equation}
The center-of-mass rapidity $\bar {y}$ of the
$\widetilde{\chi}_{i}^{0} \widetilde{\chi}_{j}^{0}$ pair, and their
respective rapidities $y_{i}^{\star}$ in their own c.m. frame, are
defined as
\begin{equation}
y_i=\bar {y}+y_{i}^{\star} ,\,\,\,\, y_j=\bar {y}+y_{j}^{\star},
\end{equation}
\begin{equation}
\Delta \equiv y_i-y_j=y_{i}^{\star}-y_{j}^{\star}.
\end{equation}
The fractional momenta of the incoming partons are expressed in the terms of their
lab-momenta by
\begin{equation}
p_1=\frac{s}{2}(x_1,0,0,x_1),\,\,\,p_2=\frac{s}{2}(x_2,0,0,-x_2),\,\,p=p_1+p_2,
\end{equation}
\begin{equation}
p^0=\frac{\sqrt s}{2}(x_1+x_2)=E_i+E_j,\,\,\,p_3=\frac{\sqrt
s}{2}(x_1-x_2)=(k_icos\theta_i+k_jcos\theta_j),
\end{equation}
which lead to
\begin{equation}
x_1=\frac{1}{2}[x_{T_i} e^{y_i}+x_{T_j} e^{y_j}]=\frac{M}{\sqrt
s}e^{\bar y},
\end{equation}
\begin{equation}
x_2=\frac{1}{2}[x_{T_i} e^{-y_i}+x_{T_j} e^{-y_j}]=\frac{M}{\sqrt
s}e^{-\bar y},
\end{equation}
\begin{equation}
\hat s=M^2=(p_1+p_2)^2=x_{1}
x_{2}s=\frac{s}{4}[x_{T_i}^2+x_{T_j}^2+2x_{T_i}x_{T_j}cosh({\Delta
y})].
\end{equation}
Using this, $\hat s$, $x_1$, $x_2$ may be calculated in terms of the
final particle rapidities $y_i$, $y_j$ and their transverse momenta.
From them, $\bar{y}$ is also obtained through ($y_{i}^{\star}$,
$y_{j}^{\star}$). The remaining Mandelstam invariants of the
subprocesses satisfy
\begin{equation}
\hat{t}=(p_1-k_i)^2=m_{\widetilde{\chi}_{i}^{0}}^2-M(E_{i}^{\star}-k^{\star}cos\theta^{\star})=
m_{\widetilde{\chi}_{i}^{0}}^2-\frac{x_{T_i}}{2}M
\sqrt{s}e^{-y_{i}^{\star}}=
\end{equation}
$$
m_{\widetilde{\chi}_{i}^{0}}^2-\frac{s}{2}x_{1} x_{T_i}e^{-y_i}=
m_{\widetilde{\chi}_{j}^{0}}^2-M(E_{j}^{\star}-k^{\star}cos\theta^{\star})=
$$
\begin{equation}
m_{\widetilde{\chi}_{j}^{0}}^2-\frac{x_{T_i}}{2}M \sqrt {s}
e^{y_{j}^{\star}}= m_{\widetilde{\chi}_{j}^{0}}^2-\frac{s}{2}x_{2}
x_{T_j}e^{y_{j}},
\end{equation}
$$
\hat{u}=(p_1-k_j)^2=m_{\widetilde{\chi}_{i}^{0}}^2-M(E_{i}^{\star}+k^{\star}cos\theta^{\star})=
m_{\widetilde{\chi}_{i}^{0}}^2-\frac{x_{T_i}}{2}M \sqrt{s}
e^{y_{i}^{\star}}=
$$
$$
m_{\widetilde{\chi}_{j}}^2-\frac{s}{2}x_{2} x_{T_i}e^{y_{i}}=
m_{\widetilde{\chi}_{j}^{0}}^2-M(E_{j}^{\star}+k^{\star}cos\theta^{\star})=
$$
\begin{equation}
m_{\widetilde{\chi}_{j}^{0}}^2-\frac{x_{T_j}}{2}M \sqrt{s}
e^{-y_{j}^{\star}}=m_{\widetilde{\chi}_{j}^{0}}^2-\frac{s}{2} x_{1}
x_{T_j} e^{-y_{j}}.
\end{equation}
\begin{equation}
\tau=\frac{\hat s}{s}=x_{1} x_{2},
\end{equation}
where $\theta^{\star}$ describes $\widetilde{\chi}_{i}^{0}$
production angle in
$\widetilde{\chi}_{i}^{0}\widetilde{\chi}_{j}^{0}$ -c.m. frame (the
$\widetilde{\chi}_{j}^{0}$ one being $\pi-\theta^{\star}$). The
energies of the two final particles in their c.m.-frame are
\begin{equation}
E_{i}^{\star}=\frac{\hat s +m_{\widetilde{\chi}_{i}^{0}}^2-
m_{\widetilde{\chi}_{j}^{0}}^2}{2 \sqrt {\hat s}},\,\,\,
E_{j}^{\star}=\frac{\hat s +m_{\widetilde{\chi}_{j}^{0}}^2-
m_{\widetilde{\chi}_{i}^{0}}^2}{2 \sqrt {\hat s}},
\end{equation}
their momentum is
\begin{equation}
p^{\star}=\frac{1}{2M}[(M^2-m_{\widetilde{\chi}_{i}^{0}}^2-m_{\widetilde{\chi}_{j}^{0}}^2)^2-
4m_{\widetilde{\chi}_{i}^{0}}^2
m_{\widetilde{\chi}_{j}^{0}}^2]^{0.5},
\end{equation}
and their velocities
\begin{equation}
\beta_{i}^{\star}={k^{\star}}/{E_{i}^{\star}}=\frac{[(M^2-m_{\widetilde{\chi}_{i}^{0}}^2-
m_{\widetilde{\chi}_{j}^{0}}^2)^2-4m_{\widetilde{\chi}_{i}^{0}}^2
m_{\widetilde{\chi}_{j}^{0}}^2]^{1/2}}
{M^2+(m_{\widetilde{\chi}_{i}}-m_{\widetilde{\chi}_{j}})^2},
\end{equation}
\begin{equation}
\beta_{j}^{\star}=k^{\star}/E_{j}^{\star}=\frac{[(M^2-m_{\widetilde{\chi}_{i}^{0}}^2-
m_{\widetilde{\chi}_{j}^{0}}^2)^2-4m_{\widetilde{\chi}_{i}^{0}}^2
m_{\widetilde{\chi}_{j}^{0}}^2]^{1/2}}
{M^2-(m_{\widetilde{\chi}_{i}}-m_{\widetilde{\chi}_{j}})^2},
\end{equation}
We also have
$$
cos\theta^{\star}=\frac{tan
y_{i}^{\star}}{{\beta}_{i}^{\star}}=-\frac{tan
y_{j}^{\star}}{{\beta}_{j}^{\star}},
$$
\begin{equation}
sin\theta^{\star}=\frac{p_T}{p^{\star}},
\end{equation}
\begin{equation}
\chi_{i}=e^{2y_{i}^{\star}}=\frac{\hat
u-m_{\widetilde{\chi}_{i}}^2}{\hat
t-m_{\widetilde{\chi}_{i}}^2}=\frac{1+\beta_{i}^{\star}cos\theta^{\star}}{1-\beta_{i}^{\star}cos\theta^{\star}},
\end{equation}
\begin{equation}
\chi_{j}=e^{2y_{j}^{\star}}=\frac{\hat
t-m_{\widetilde{\chi}_{j}}^2}{\hat
u-m_{\widetilde{\chi}_{j}}^2}=\frac{1-\beta_{j}^{\star}cos\theta^{\star}}{1+\beta_{j}^{\star}cos\theta^{\star}},
\end{equation}
\begin{equation}
\beta_{i}^{\star}cos\theta^{\star}=\frac{\hat u-\hat t}{\hat u+\hat
t}=\frac{{\chi}_{i}-1}{{\chi}_{i}+1},
\end{equation}
\begin{equation}
\chi_{j}=\frac{{\chi}_{i}(m_{\widetilde{\chi}_{j}}^2-m_{\widetilde{\chi}_{i}}^2)+M^2}
{{\chi}_{i}M^2+m_{\widetilde{\chi}_{j}}^2-m_{\widetilde{\chi}_{i}}^2},
\end{equation}
\begin{equation}
E_{T_{i}}=\frac{E_{i}^{\star}}{coshy_{i}^{\star}},
\end{equation}
\begin{equation}
k_{T}^2=\frac{(M^2+m_{\widetilde{\chi}_{i}^{0}}^2-m_{\widetilde{\chi}_{j}^{0}}^2)^2{\chi}_{i}-M^2
m_{\widetilde{\chi}_{i}^{0}}^2(1+{\chi}_{i})^2}{M^2(1+{\chi}_{i})^2},
\end{equation}
\begin{equation}
x_{T_{i}}^2=\frac{4(M^2+m_{{\chi}_{i}^{0}}^2-m_{{\chi}_{j}^{0}})^2{\chi}_{i}}{M^2
s(1+{\chi}_{i})^2},
\end{equation}
\begin{equation}
x_{T_{j}}^2=\frac{4(M^2+m_{{\chi}_{j}}^2-m_{{\chi}_{i}})^2{\chi}_{j}}{M^2
s(1+{\chi}_{j})^2}.
\end{equation}
Using (3.20) we define the  expression for the cross section in
terms of the overall center-of-mass rapidities of the two jets
yielding
\begin{equation}
\frac{d\sigma}{dy_{i}dy_{j}dk_{T}^2}=x_{1}x_{2}\sum_{q}G_{{q_{1}}/{h_{1}}}(x_{1},Q)
 G_{{q_{2}}/{h_{2}}}(x_{2},Q)\frac{d\sigma}{d\hat t}(q_1 q_2 \to
\widetilde{\chi}_{i}^{0}\widetilde{\chi}_{j}^{0}).
\end{equation}
\section{Numerical results and discussion}\label{results}

In this section, we discuss the numerical results for the process
$pp \to \widetilde{\chi}_{i}^{0}\widetilde{\chi}_{j}^{0}$ at LHC
(c.m energy $\sqrt s$=14 TeV) generated by the subprocesses $q
\bar{q}$. As we assume, $\widetilde{\chi}_{1}^{0}$ is likely to be
the LSP, the three types of channels: $q \bar{q} \to \widetilde
{\chi}_{1}^{0}\widetilde{\chi}_{1}^{0}$, $q \bar{q} \to \widetilde
{\chi}_{2}^{0}\widetilde{\chi}_{2}^{0}$, $q \bar{q} \to \widetilde
{\chi}_{1}^{0}\widetilde{\chi}_{2}^{0}$, would be the most dominant
neutralino pair production processes, which may lead to the first
detection of SUSY particles at the LHC. The numerical results of the
cross sections of these three processes have been presented and
their dependencies on the basic SUSY parameters have been discussed.
We divide the input MSSM parameters into two parts. One part is for
the general parameters included also in SM, and the other part is
the ino and squark sectors of MSSM. For the first parameter part, we
take $m_{Z^{0}}$= 91.1887 GeV, $\sin\Theta_{W}^2=0.2315$,
$\alpha_{E}$=1/137. For the second part, we just limit the values of
$M_1$, $M_2$ and $\mu$ to be real, positive and below 1 TeV, and
take $\tan\beta=2, m_{\widetilde{u}_{1}}=m_{\widetilde{d}_{1}}$=300
GeV, $m_{\widetilde{u}_{2}}=m_{\widetilde{d}_{2}} $=500 GeV. Also,
we fix the heavy chargino mass as
$m_{\widetilde{\chi}_{2}^{+}}=450GeV$ and for the lightest chargino
mass $m_{\widetilde{\chi}_{1}^{+}}=150GeV$. By using Eqs.(2.7, 2.8)
with above chargino mass values, one may have two choices of
parameter sets for $\mu$ and $M_2$ in two extreme cases, which are
the Higgsino-like and the gaugino-like respectively. For the
Higgsino-like case, we get $M_2=437.96 GeV$, $\mu=169.753 GeV$,
$M_1=219.703 GeV$  and by inserting the values of $M_2$, $\mu$ and
$M_1$ into Eq.(2.11) for neutralino masses, we get
$$
m_{\widetilde{\chi}_{1}^{0}}=123.242 GeV, \,\,
m_{\widetilde{\chi}_{2}^{0}}=179.475 GeV, \,\,
m_{\widetilde{\chi}_{3}^{0}}=238.199 GeV, \,\,
m_{\widetilde{\chi}_{4}^{0}}=458.479 GeV \
$$
For the gaugino-like case, we have $M_2=169.753GeV$,
$\mu=437.96GeV$, $M_1=85.157 GeV$ and by inserting the values of
$M_2$, $\mu$ and $M_1$ into Eq.(2.11) for neutralino masses, we then
get
$$
m_{\widetilde{\chi}_{1}^{0}}=77.212 GeV, \,\,
m_{\widetilde{\chi}_{2}^{0}}=153.859 GeV, \,\,
m_{\widetilde{\chi}_{3}^{0}}=449.420 GeV, \,\,
m_{\widetilde{\chi}_{4}^{0}}=452.443 GeV
$$
For full discussion, we present the results for the mixture case as
well. We take the heavy chargino mass
$m_{\widetilde{\chi}_{2}^{+}}=280.6 GeV$ and for the lightest
chargino mass $m_{\widetilde{\chi}_{1}^{+}}=128 GeV,$ the
corresponding outputs obtained as $M_2= \mu=200.598 GeV$,
$M_1=100.63 GeV$ and also by inserting the values of $M_2$, $\mu$
and $M_1$ into Eq.(2.11) for the neutralino masses, we get
$$
m_{\widetilde{\chi}_{1}^{0}}=72.576 GeV, \,\,
m_{\widetilde{\chi}_{2}^{0}}=147.127 GeV, \,\,
m_{\widetilde{\chi}_{3}^{0}}=207.184 GeV, \,\,
m_{\widetilde{\chi}_{4}^{0}}=278.635 GeV
$$
As an example for the quark distribution function inside the
proton, we use the MRST2003c package[17]. After this, the cross
sections of the subprocesses $q \bar{q} \to \widetilde
{\chi}_{1}^{0}\widetilde{\chi}_{1}^{0}$,
$\widetilde{\chi}_{1}^{0}\widetilde{\chi}_{2}^{0}$,
$\widetilde{\chi}_{2}^{0}\widetilde{\chi}_{2}^{0}$ can be
numerically evaluated. For illustration, we have calculated the
total cross sections of the subprocesses $u \bar{u} \to \widetilde
{\chi}_{1}^{0}\widetilde{\chi}_{1}^{0}$,
$\widetilde{\chi}_{1}^{0}\widetilde{\chi}_{2}^{0}$,
$\widetilde{\chi}_{2}^{0}\widetilde{\chi}_{2}^{0}$, and $d \bar{d}
\to \widetilde {\chi}_{1}^{0}\widetilde{\chi}_{1}^{0}$,
$\widetilde{\chi}_{1}^{0}\widetilde{\chi}_{2}^{0}$,
$\widetilde{\chi}_{2}^{0}\widetilde{\chi}_{2}^{0}$, the dependence
on the beam energies, on the mass of squarks and also of the $M_2$
gaugino mass. We will illustrate this for three extremely
different scenarios. In Fig.2-4, we show the dependence of the
total cross sections for the subprocesses $u \bar{u} \to
\widetilde {\chi}_{1}^{0}\widetilde{\chi}_{1}^{0}$,
$\widetilde{\chi}_{1}^{0}\widetilde{\chi}_{2}^{0}$,
$\widetilde{\chi}_{2}^{0}\widetilde{\chi}_{2}^{0}$ of the beam
energy. In our calculations, the beam energy for the subprocesses
$\hat s$ changes in the region $(400\div5000)GeV$, which
corresponds to the beam energy for process $s$
$(500\div16000)GeV$. As shown in Fig.2-4, all subprocesses in the
gaugino-like scenario, the total cross section is 30 percent
larger than the mixing scenario in magnitude, and larger than in
the Higgsino-like scenario as 1.79-2 order of magnitude. As seen
from Figs.2-4, in all three scenarios, the dependencies of the
cross section on the beam energy demonstrate the same behavior.
Also, as seen from Figs.2-4, all three scenarios, by increasing
the beam energy from 400 GeV to 3600 GeV, the total cross section
is monotonically increasing. In Figs.5-7, we show the dependence
of the total cross sections for the subprocesses $d \bar{d} \to
\widetilde {\chi}_{1}^{0}\widetilde{\chi}_{1}^{0}$,
$\widetilde{\chi}_{1}^{0}\widetilde{\chi}_{2}^{0}$,
$\widetilde{\chi}_{2}^{0}\widetilde{\chi}_{2}^{0}$ of the beam
energy. As shown in Fig.5-6, all subprocesses in the gaugino-like
scenario, the total cross section is 8 percent larger than the
mixing scenario in magnitude, and larger than in the Higgsino like
scenario as 2-2.137 order of magnitude. As shown in Fig.7, in the
gaugino-like scenario, the total cross section is almost 2 percent
larger than the mixing scenario in magnitude, and larger than in
the Higgsino-like scenario as two order of magnitude. As seen from
Figs.5-7, in all three scenarios, the dependence of the cross
section on the beam energy demonstrates the same behavior. With
the increase of the beam energy from 400 GeV to 4000 GeV, the
total cross section is monotonically increasing. The total cross
section of the subprocesses $u \bar{u} \to \widetilde
{\chi}_{1}^{0}\widetilde{\chi}_{2}^{0}$ in the gaugino-like
scenario, appears in the range of 2 to 140 fb and should be
observable at LHC. It should be noted that, one of the
subprocesses, $q \bar{q} \to \widetilde
{\chi}_{1}^{0}\widetilde{\chi}_{2}^{0}$ dominates all three
scenarios. According to our opinion, it may be used as a probe for
an experimental search on the neutralino pair. Figs.8-13 show the
dependence of the total cross sections for the subprocesses $u
\bar{u} \to \widetilde {\chi}_{1}^{0}\widetilde{\chi}_{1}^{0}$,
$\widetilde{\chi}_{1}^{0}\widetilde{\chi}_{2}^{0}$,
$\widetilde{\chi}_{2}^{0}\widetilde{\chi}_{2}^{0}$, and $d \bar{d}
\to \widetilde {\chi}_{1}^{0}\widetilde{\chi}_{1}^{0}$,
$\widetilde{\chi}_{1}^{0}\widetilde{\chi}_{2}^{0}$,
$\widetilde{\chi}_{2}^{0}\widetilde{\chi}_{2}^{0}$, of the
$M_{2}$. As seen from Figs.8-10, the cross section is decreasing
when the $M_2$ gaugino mass is increasing. But, as seen from
Fig.11, the dependence of the total cross sections for the
subprocesses $d \bar{d} \to \widetilde
{\chi}_{1}^{0}\widetilde{\chi}_{1}^{0}$,
$\widetilde{\chi}_{1}^{0}\widetilde{\chi}_{2}^{0}$,
$\widetilde{\chi}_{2}^{0}\widetilde{\chi}_{2}^{0}$ of the $M_2$
have another character. As shown in Figs.12-13, the cross section
is decreasing with increasing $M_2$ in the range of 300 to 400 and
it has a minimum approximately at one point $M_2=400 GeV$. After
this point, the cross section increases with $M_2$ gaugino mass
increasing. Behavior of the cross section is decreasing when the
$M_2$ gaugino mass is increasing as expected. Because, we have
obtained the following relation for the dependence cross section
depending on the beam energy:
$\sigma(Higgsino-like)<\sigma(Mixture-case)<\sigma(Gaugino-like)$.
This relation  corresponds to
$M_2(Gaugino-like)<M_2(Mixture-case)<M_2(Higgsino-like)$,
respectively. Therefore, the cross section is decreasing when the
$M_2$ gaugino  mass is increasing. Figs.14-19 show the dependence
of the total cross sections for the subprocesses $u\bar{u} \to
\widetilde {\chi}_{1}^{0}\widetilde{\chi}_{1}^{0}$,
$\widetilde{\chi}_{1}^{0}\widetilde{\chi}_{2}^{0}$,
$\widetilde{\chi}_{2}^{0}\widetilde{\chi}_{2}^{0}$, and $d \bar{d}
\to \widetilde {\chi}_{1}^{0}\widetilde{\chi}_{1}^{0}$,
$\widetilde{\chi}_{1}^{0}\widetilde{\chi}_{2}^{0}$,
$\widetilde{\chi}_{2}^{0}\widetilde{\chi}_{2}^{0}$, of the squarks
mass. As shown in Figs.14-19, in all subprocesses and scenarios,
the cross section monotonically decreases with an increase in the
squark masses. Therefore, an experimental search for the
neutralino pair at lower values of the squark mass is preferable.
In Figs.20-22, we show the dependence of the differential cross
sections for the process $pp \to
\widetilde{\chi}_{i}^0\widetilde{\chi}_{j}^0$  as a function of
the $k_T$ transverse momentum of the neutralino pair at rapidity
$y_i=y_j=0$. As seen from Figs.20-22, the differential cross
sections decrease monotonically pursuant to an increase in the
$k_T$ transverse momentum of the neutralino pair. It should be
noted that, one of the process $pp \to
\widetilde{\chi}_{1}^0\widetilde{\chi}_{2}^0$ dominates all three
scenarios. As mentioned in above, that our results disagree with
the results in [6]. So, the expression for the cross sections in
Ref.[6] is wrong. Therefore, our  numerical results also disagree
with the numerical results in [6]. We compared our results with
[18]. The pair production rate of
$\widetilde{\chi}_{1}^0\widetilde{\chi}_{2}^0$ via gluon-gluon
fusion is about few ten percent of that via quark-antiquark
annihilation at the LHC. The cross section of
$\widetilde{\chi}_{2}^0\widetilde{\chi}_{2}^0$ via gluon-gluon
fusion is about few  percent of that via quark-antiquark
annihilation at the LHC.

\section{Conclusion}\label{Conc}
In this paper, we have considered the neutralino pair production
processes in proton-proton collisions at LHC (c.m energy $\sqrt s=14
TeV$). In the description, we have taken into account the subprocess
$ q(p_1)\bar q(p_2) \to
\widetilde\chi_{i}^{0}(k_1)\widetilde\chi_{j}^{0}(k_2)$. We have
given detail illustrations for the center-of-mass energy, squarks
and $M_2$ gaugino masses for three extremely different scenarios.
For illustration, we have calculated the total cross sections of the
subprocesses $q \bar{q} \to \widetilde
{\chi}_{1}^{0}\widetilde{\chi}_{1}^{0}$,
$\widetilde{\chi}_{1}^{0}\widetilde{\chi}_{2}^{0}$,
$\widetilde{\chi}_{2}^{0}\widetilde{\chi}_{2}^{0}$, and the
dependence on the beam energies, on the mass of squarks and also of
the $M_2$ gaugino mass. We have illustrated this for three extremely
different scenarios. Fig.2-7 show the dependence of the total cross
sections for the subprocesses $q \bar{q} \to \widetilde
{\chi}_{1}^{0}\widetilde{\chi}_{1}^{0}$,
$\widetilde{\chi}_{1}^{0}\widetilde{\chi}_{2}^{0}$,
$\widetilde{\chi}_{2}^{0}\widetilde{\chi}_{2}^{0}$ of the beam
energy. As seen from Figs.2-7, in all three scenarios, the
dependencies of the cross section on the beam energy demonstrate the
same behavior. The total cross section of the subprocesses $u
\bar{u} \to \widetilde {\chi}_{1}^{0}\widetilde{\chi}_{2}^{0}$ in
the gaugino-like scenario, appears in the range of 2 to 140 fb and
should be observable at LHC. It should be noted that, one of the
subprocesses, $u \bar{u} \to \widetilde
{\chi}_{1}^{0}\widetilde{\chi}_{2}^{0}$ dominates all three
scenarios. According to our opinion, it may be used as a probe for
an experimental search on the neutralino pair. Figs.8-13 show the
dependence of the total cross sections for the subprocesses $q
\bar{q} \to \widetilde {\chi}_{1}^{0}\widetilde{\chi}_{1}^{0}$,
$\widetilde{\chi}_{1}^{0}\widetilde{\chi}_{2}^{0}$,
$\widetilde{\chi}_{2}^{0}\widetilde{\chi}_{2}^{0}$, of the $M_{2}$.
Figs.14-19 show the dependence of the total cross sections for the
subprocesses $q \bar{q} \to \widetilde
{\chi}_{1}^{0}\widetilde{\chi}_{1}^{0}$,
$\widetilde{\chi}_{1}^{0}\widetilde{\chi}_{2}^{0}$,
$\widetilde{\chi}_{2}^{0}\widetilde{\chi}_{2}^{0}$ of the squarks
mass. As shown in Figs.14-19, in all subprocesses and scenarios, the
cross section monotonically decreases with an increase in squark
masses. It should be underlined that for all three scenarios, the
dependence of the total cross section of the subprocesses
$q\bar{q}\to\widetilde{\chi}_{i}^{0}\widetilde{\chi}_{j}^0$ on the
beam energy is dominated by one of the subprocesses,
$q\bar{q}\to\widetilde{\chi}_{1}^{0}\widetilde{\chi}_{2}^0$ and the
dependence of the total cross section of the subprocesses
$q\bar{q}\to\widetilde{\chi}_{i}^{0}\widetilde{\chi}_{j}^0$ on the
mass of squarks is dominated by one of the subprocesses,
$q\bar{q}\to\widetilde{\chi}_{1}^{0}\widetilde{\chi}_{1}^0$. These
findings may be used as a probe in an experimental search for the
neutralino pair at LHC. In Figs.20-22, we have shown the dependence
of the differential cross sections for the process $pp \to
\widetilde{\chi}_{i}^0\widetilde{\chi}_{j}^0$  as a function of the
$k_T$ transverse momentum of the neutralino pair at rapidity
$y_i=y_j=0$. In this case, one of the process $pp \to
\widetilde{\chi}_{1}^0\widetilde{\chi}_{2}^0$ dominate all three
scenarios. These features make the neutralino pair production
processes rather interesting for testing the SUSY dynamics at LHC.
The reason is that they provide tests which will be complementary to
those addressing the cascade decays of initially produced colored
SUSY particles to eventually $\widetilde{\chi}_{1}^0$, which is here
assumed to be the LSP; e.g. studies of mass spectra and decay
branching ratios [19]. In particular, consistency checks should thus
become available, allowing the strengthening of possible constraints
on the validity of specific models. Moreover, in such neutralino
pair production, the role of the Majorana nature of the final state
particles is more prominent than in decays involving just one
neutralino at a time. Since no such states have been observed in the
past, it would be interesting to have eventually some experimental
support of our understanding of the Majorana nature. These results
imply an interesting complementarity between the future LHC
measurements, the related $\gamma\gamma \to
\widetilde\chi_{i}^{0}\widetilde\chi_{j}^{0}$ measurements at a
future Linear Collider and the Dark Matter searches in cosmic
experiments.
\section*{Acknowledgments}
Two of authors, A.~Ahmadov and I.~Boztosun are grateful to T\"{U}B\.{I}TAK
Grant-2221 (BAYG) as well as T\"{U}B\.{I}TAK Grant: TBAG-2398. One of us, A.~Ahmadov is grateful also to NATO
Reintegration Grant-980779

\begin{figure}[htb]
     \begin{center}
\mbox{\epsfig{figure=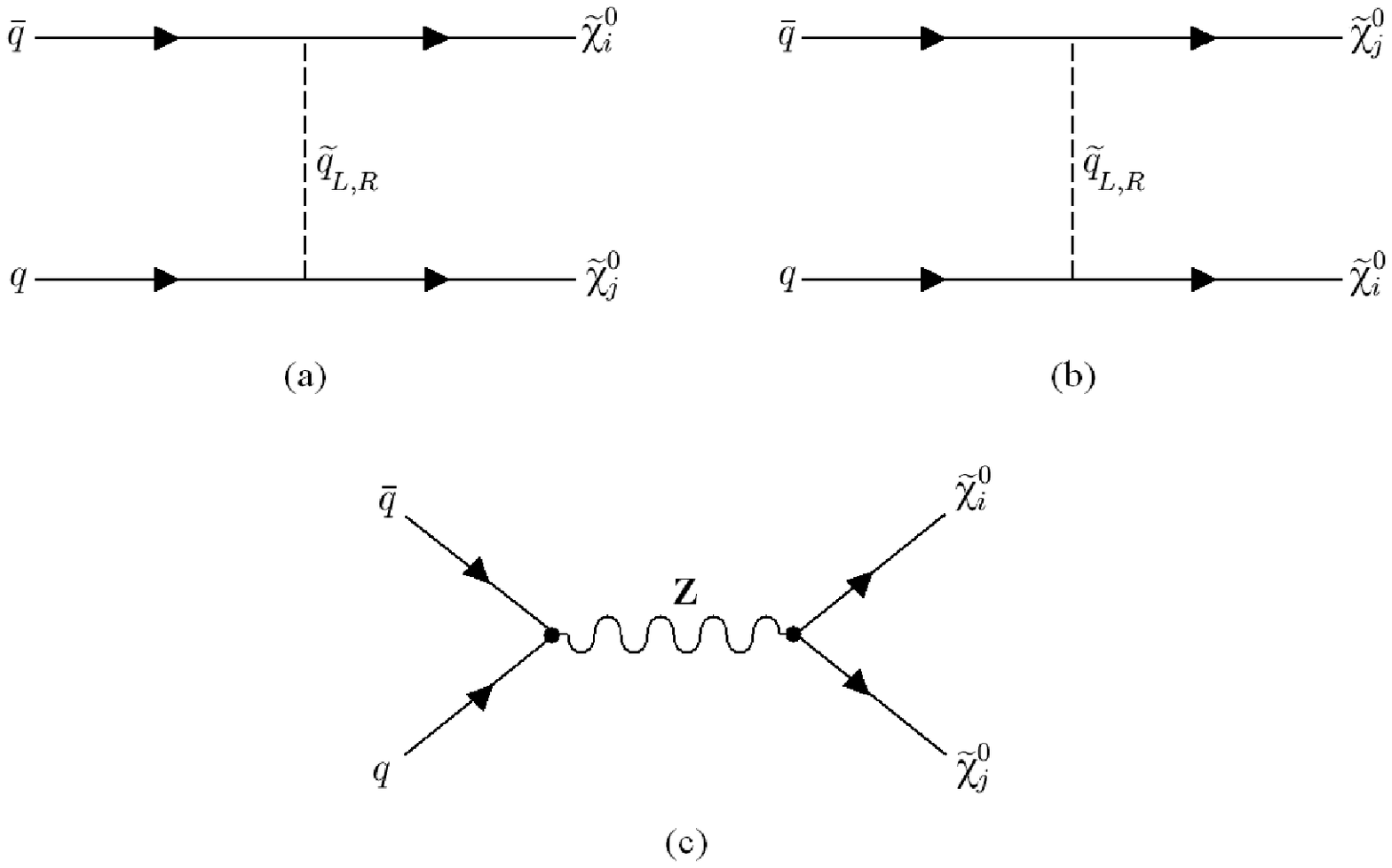
                             ,height=20cm}}
     \end{center}
\caption{Feynman diagrams for $q\bar{q}
\to\widetilde{\chi}_{i}^{0}\widetilde{\chi}_{j}^{0}$ process.}
\label{Fig1}
\end{figure}

\newpage

\begin{figure}[htb]
    \begin{center}
\mbox{\epsfig{figure=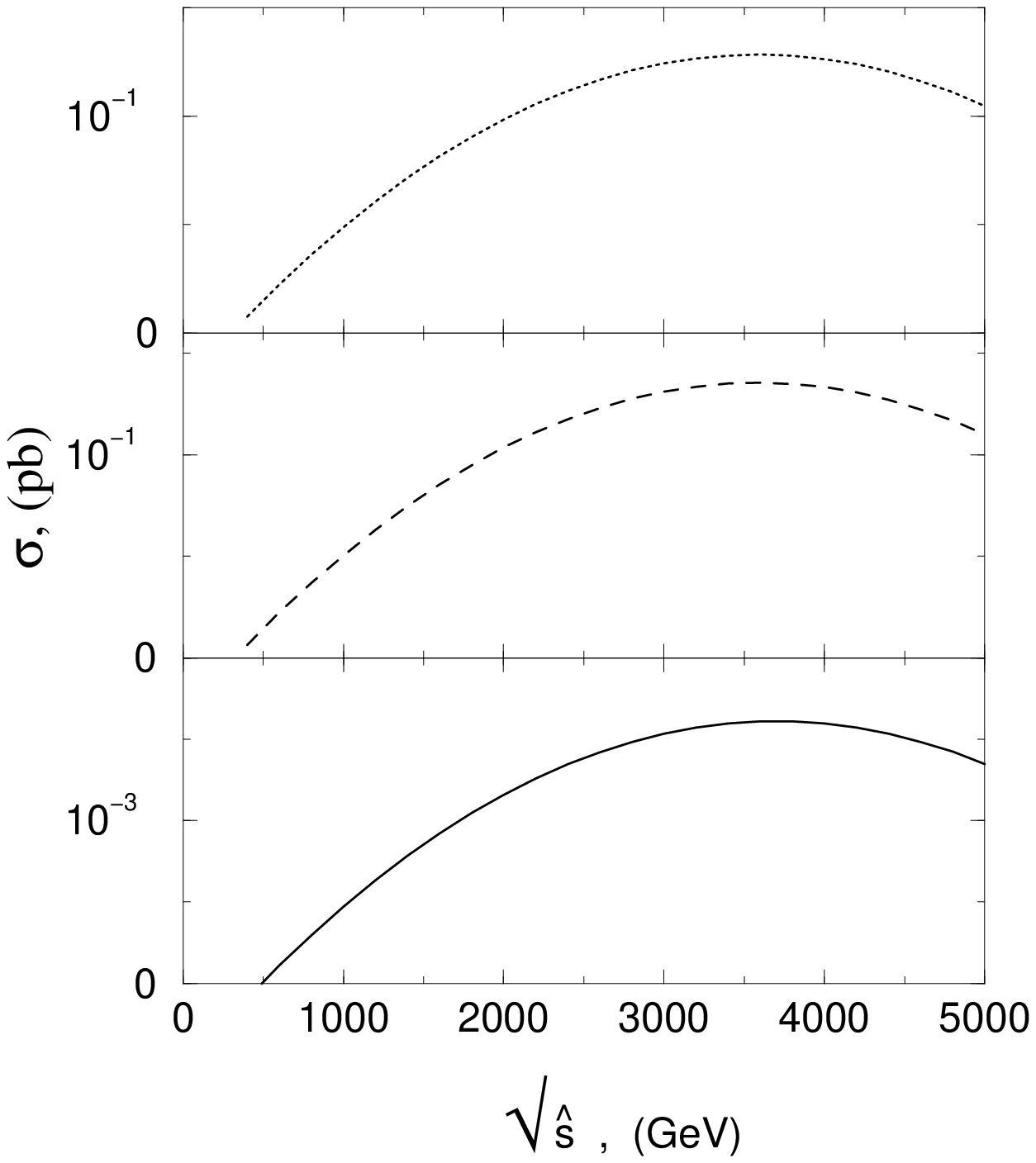
                             ,height=7.3cm,width=8.1cm}}
     \end{center}
\caption{The cross sections of the subprocess
$u\bar{u}\to\widetilde{\chi}_{1}^0\widetilde{\chi}_{1}^0$ as a
function of $\sqrt{\hat{s}}$. The curves correspond to:
solid-Higgsino-like, dashed-gaugino-like and dotted-mixture cases,
respectively.}
\label{Fig2}
\end{figure}

\begin{figure}[htb]
    \begin{center}
\mbox{\epsfig{figure=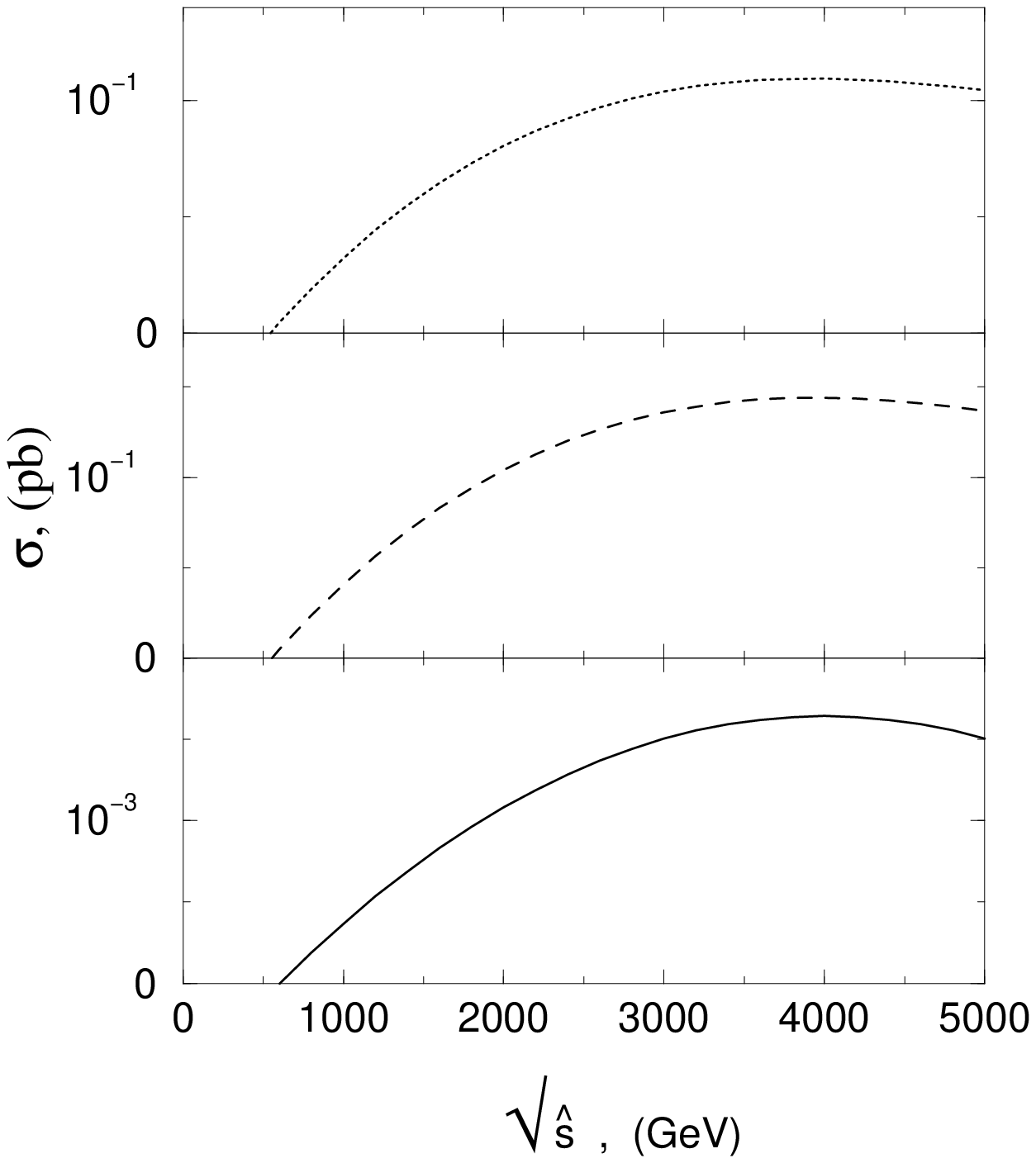
                            ,height=7.3cm,width=8.1cm}}
     \end{center}
\caption{The cross sections of the subprocess
$u\bar{u}\to\widetilde{\chi}_{1}^0\widetilde{\chi}_{2}^0$ as a
function of $\sqrt{\hat{s}}$. The curves correspond to:
solid-Higgsino-like, dashed-gaugino-like and dotted-mixture cases,
respectively.}
\label{Fig3}
\end{figure}

\newpage

\begin{figure}[htb]
    \begin{center}
\mbox{\epsfig{figure=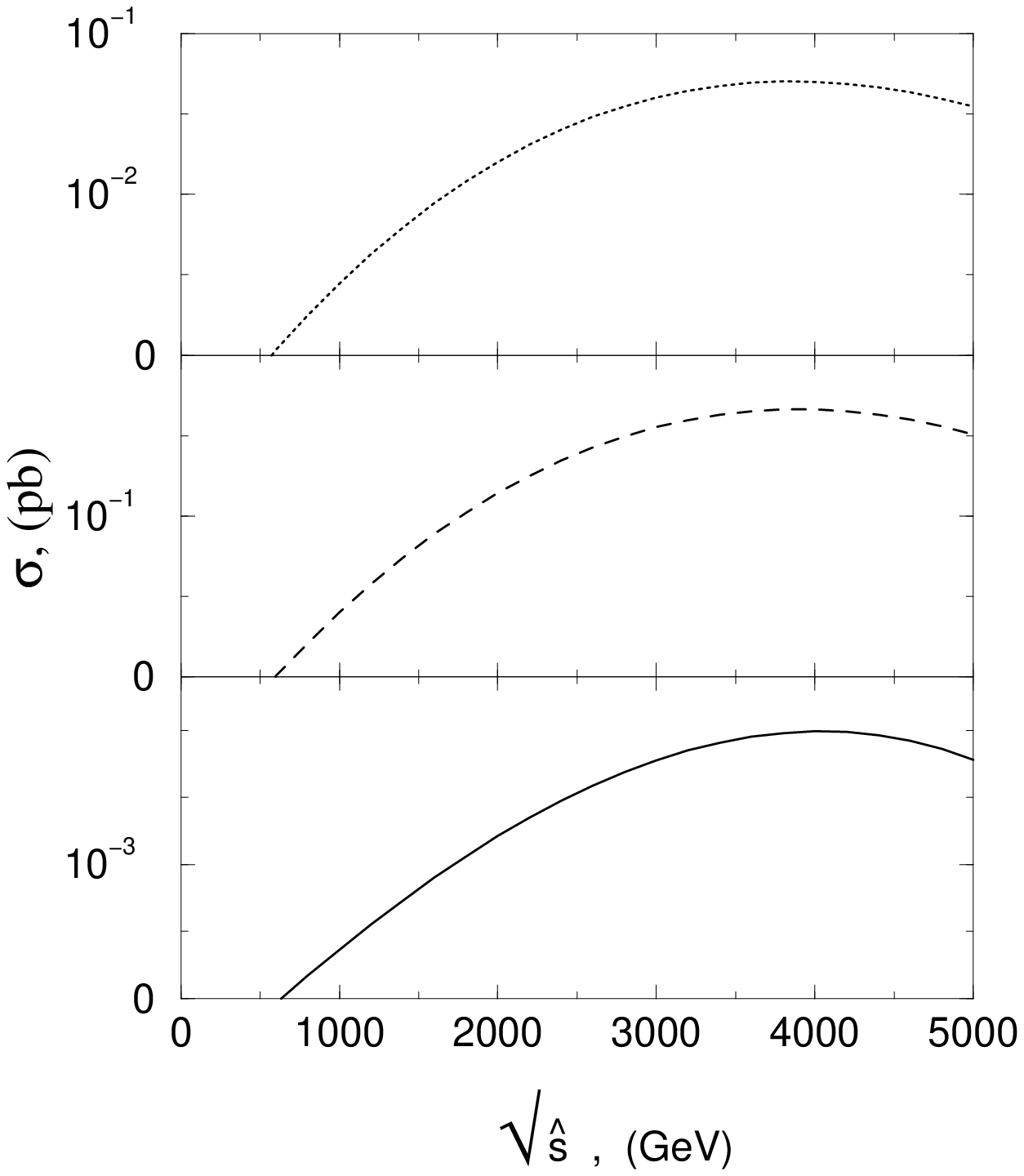
                             ,height=7.3cm,width=8.1cm}}
      \end{center}
\caption{The cross sections of the subprocess
$u\bar{u}\to\widetilde{\chi}_{2}^0\widetilde{\chi}_{2}^0$ as a
function of $\sqrt{\hat{s}}$. The curves correspond to:
solid-Higgsino-like, dashed-gaugino-like and dotted-mixture cases,
respectively.}
\label{Fig4}
\end{figure}

\begin{figure}[htb]
    \begin{center}
\mbox{\epsfig{figure=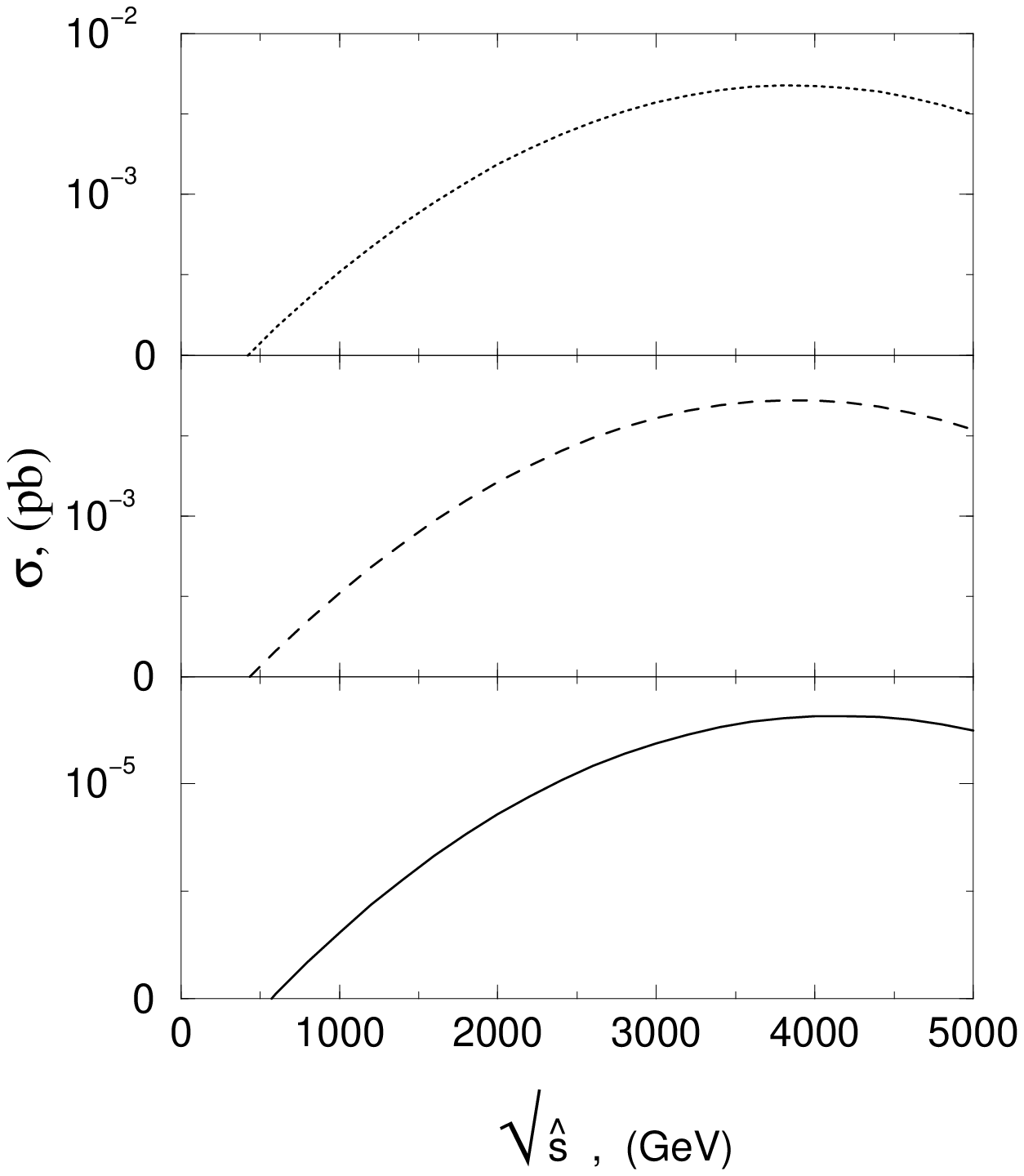
                             ,height=7.2cm,width=7.8cm}}
      \end{center}
\caption{The cross sections of the subprocess
$d\bar{d}\to\widetilde{\chi}_{1}^0\widetilde{\chi}_{1}^0$ as a
function of $\sqrt{\hat{s}}$. The curves correspond to:
solid-Higgsino-like, dashed-gaugino-like and dotted-mixture cases,
respectively.} \label{Fig5}
\end{figure}

\newpage

\begin{figure}[htb]
    \begin{center}
\mbox{\epsfig{figure=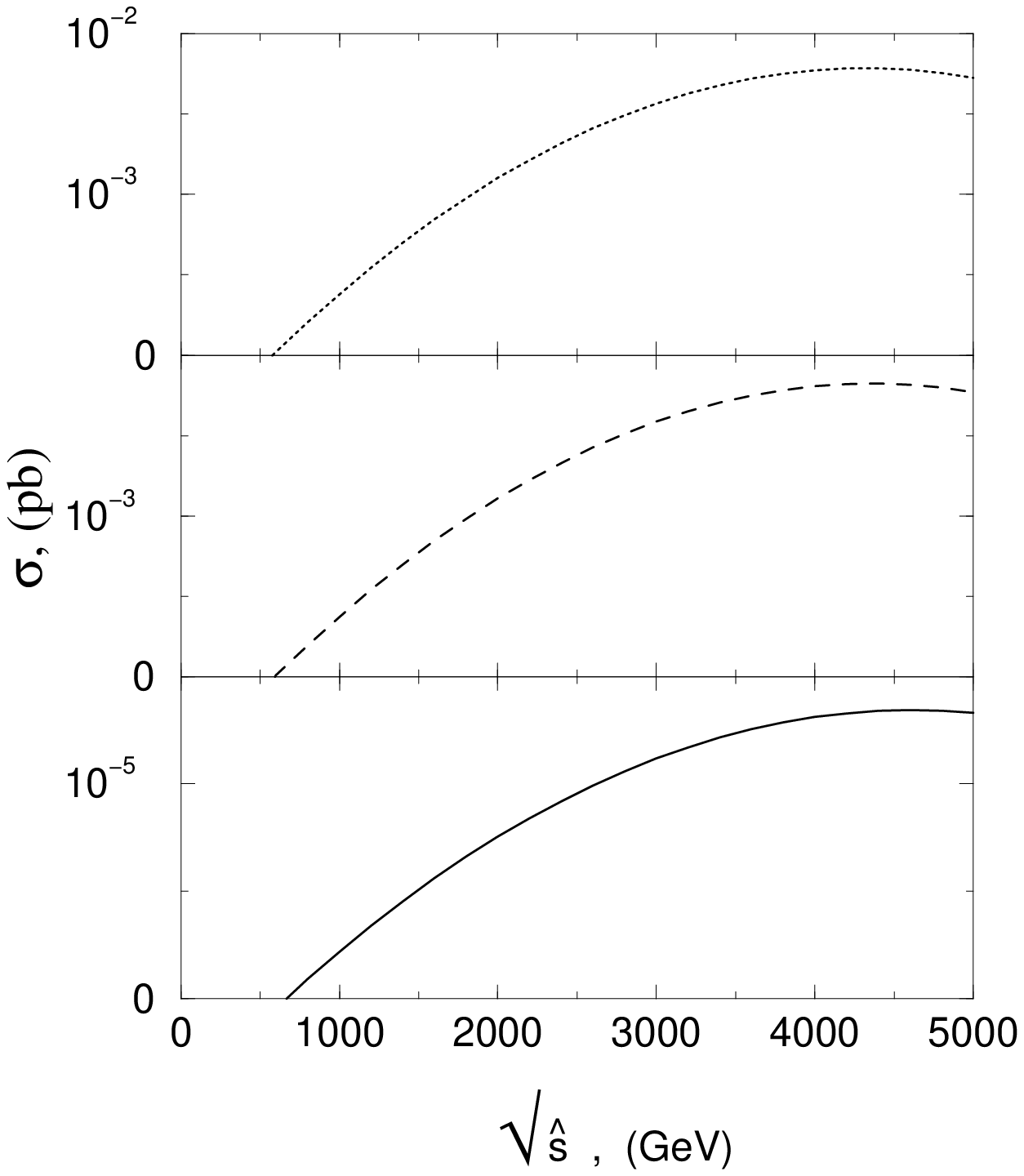
                             ,height=7.2cm,width=7.8cm}}
      \end{center}
\caption{The cross sections of the subprocess
$d\bar{d}\to\widetilde{\chi}_{1}^0\widetilde{\chi}_{2}^0$ as a
function of $\sqrt{\hat{s}}$. The curves correspond to:
solid-Higgsino-like, dashed-gaugino-like and dotted-mixture cases,
respectively.} \label{Fig6}
\end{figure}

\begin{figure}[htb]
    \begin{center}
\mbox{\epsfig{figure=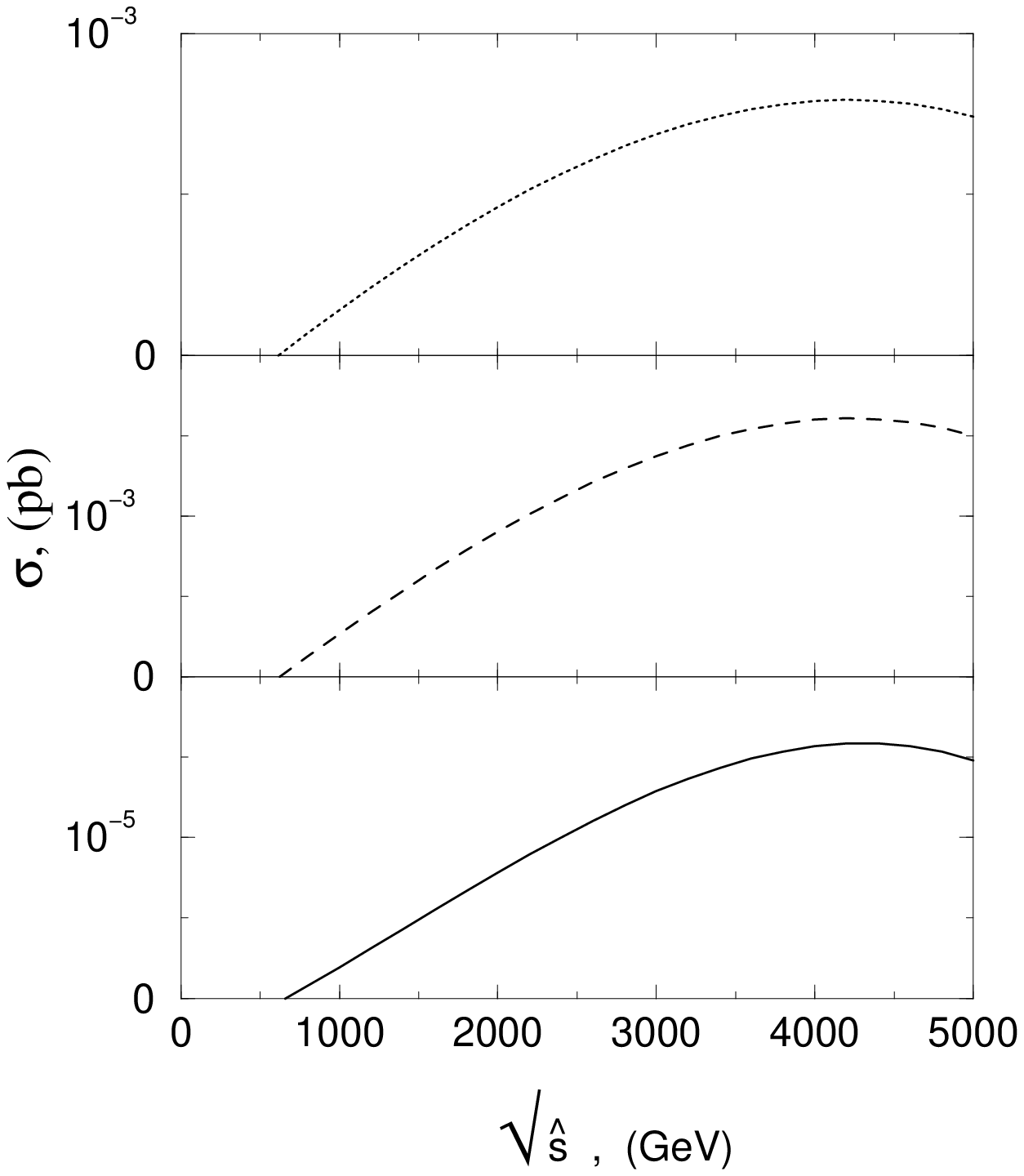
                             ,height=7.2cm,width=7.9cm}}
      \end{center}
\caption{The cross sections of the subprocess
$d\bar{d}\to\widetilde{\chi}_{2}^0\widetilde{\chi}_{2}^0$ as a
function of $\sqrt{\hat{s}}$. The curves correspond to:
solid-Higgsino-like, dashed-gaugino-like and dotted-mixture cases,
respectively.} \label{Fig7}
\end{figure}

\newpage

\begin{figure}[htb]
    \begin{center}
\mbox{\epsfig{figure=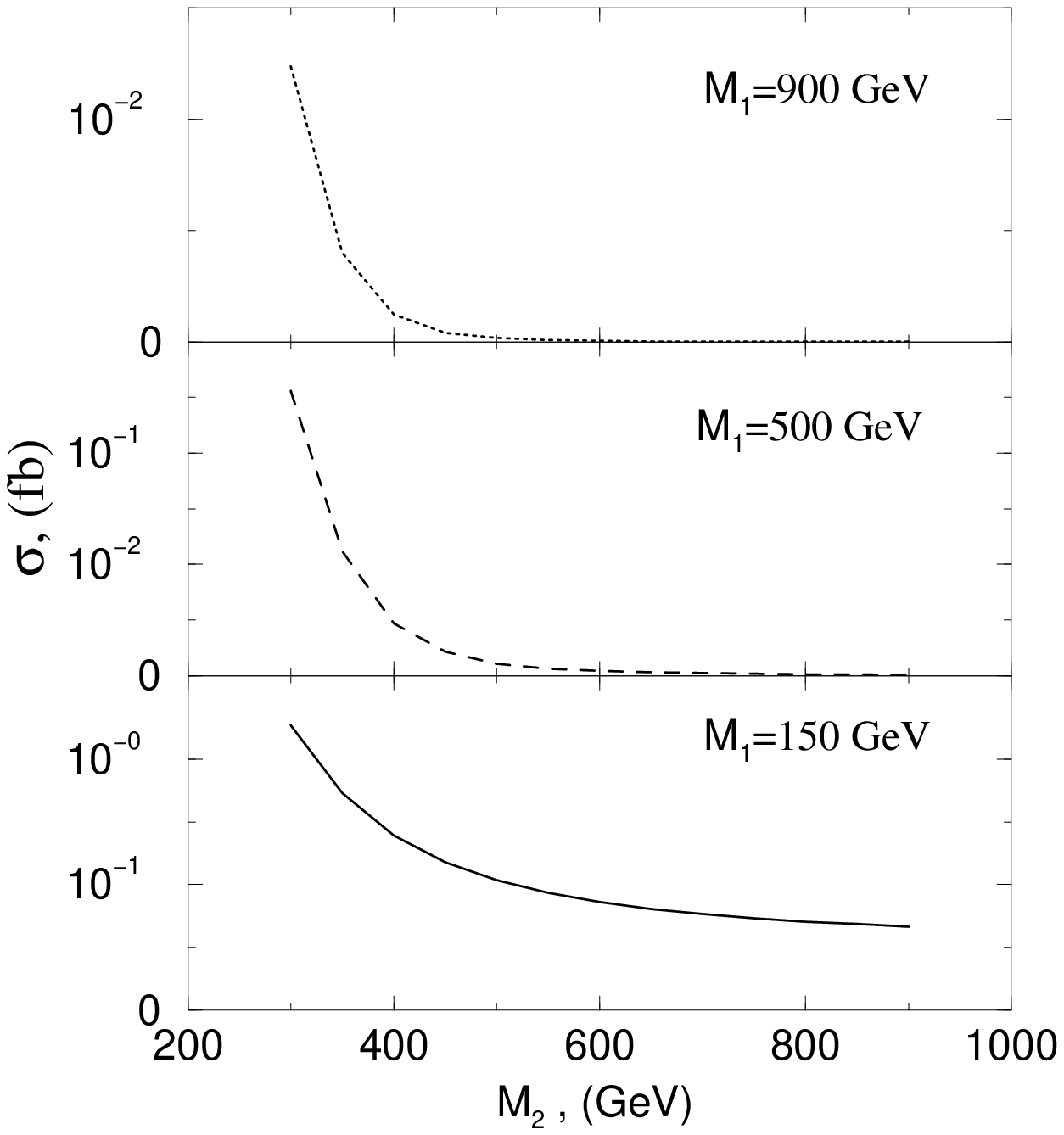
                             ,height=7.2cm,width=7.9cm}}
      \end{center}
\caption{The cross sections of the subprocess
$u\bar{u}\to\widetilde{\chi}_{1}^{0}\widetilde{\chi}_{1}^0$, as a
function of $M_2$ with $\mu = 450 $ GeV and $\sqrt{\hat{s}} =1.5$
TeV}
\label{Fig8}
\end{figure}

\begin{figure}[htb]
    \begin{center}
\mbox{\epsfig{figure=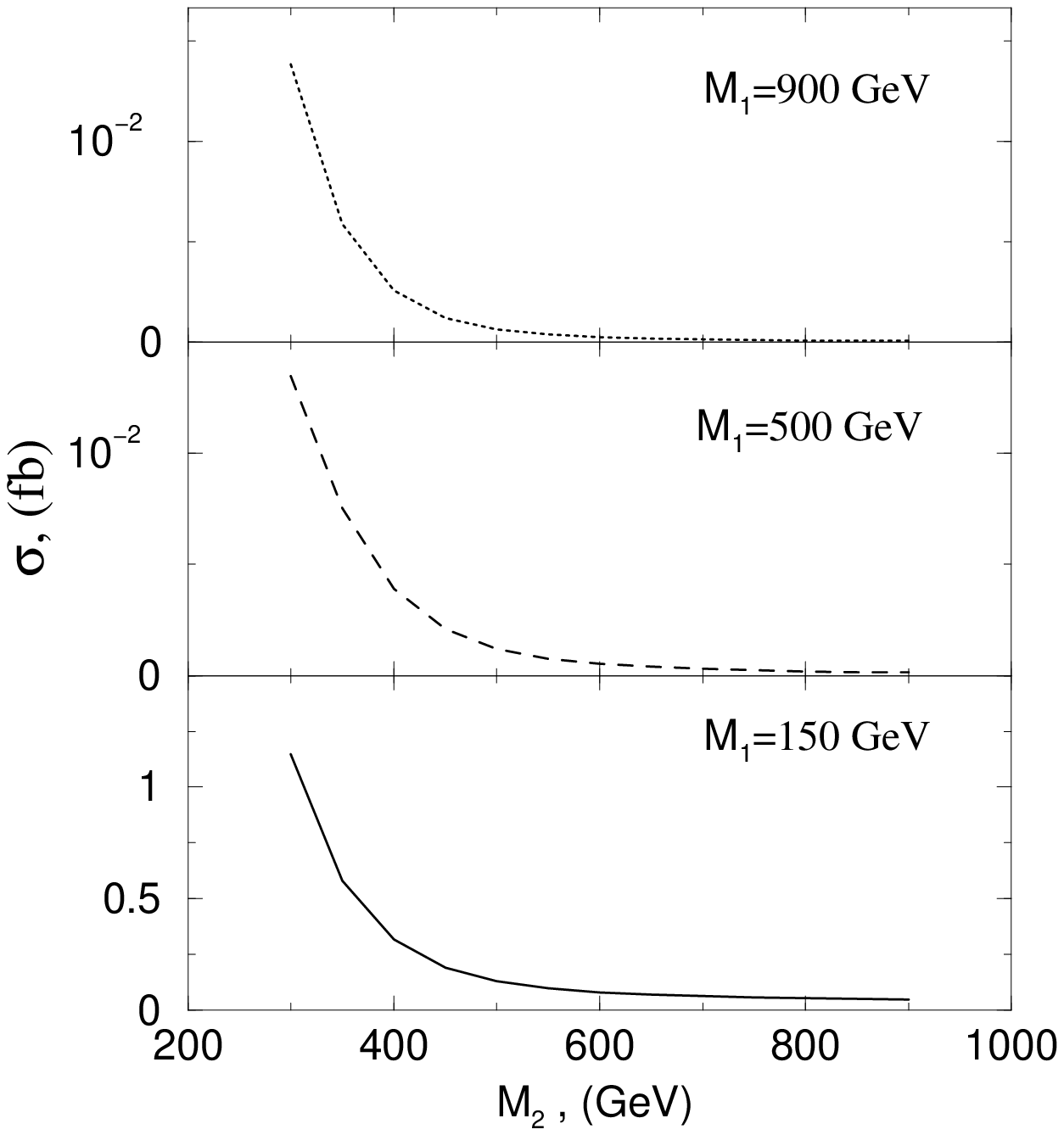
                             ,height=7.1cm,width=7.5cm}}
      \end{center}
\caption{The cross sections of the subprocess
$u\bar{u}\to\widetilde{\chi}_{1}^{0}\widetilde{\chi}_{2}^0$, as a
function of $M_2$ with $\mu = 450 $ GeV and $\sqrt{\hat{s}} =1.5$
TeV}
\label{Fig9}
\end{figure}

\newpage

\begin{figure}[htb]
    \begin{center}
\mbox{\epsfig{figure=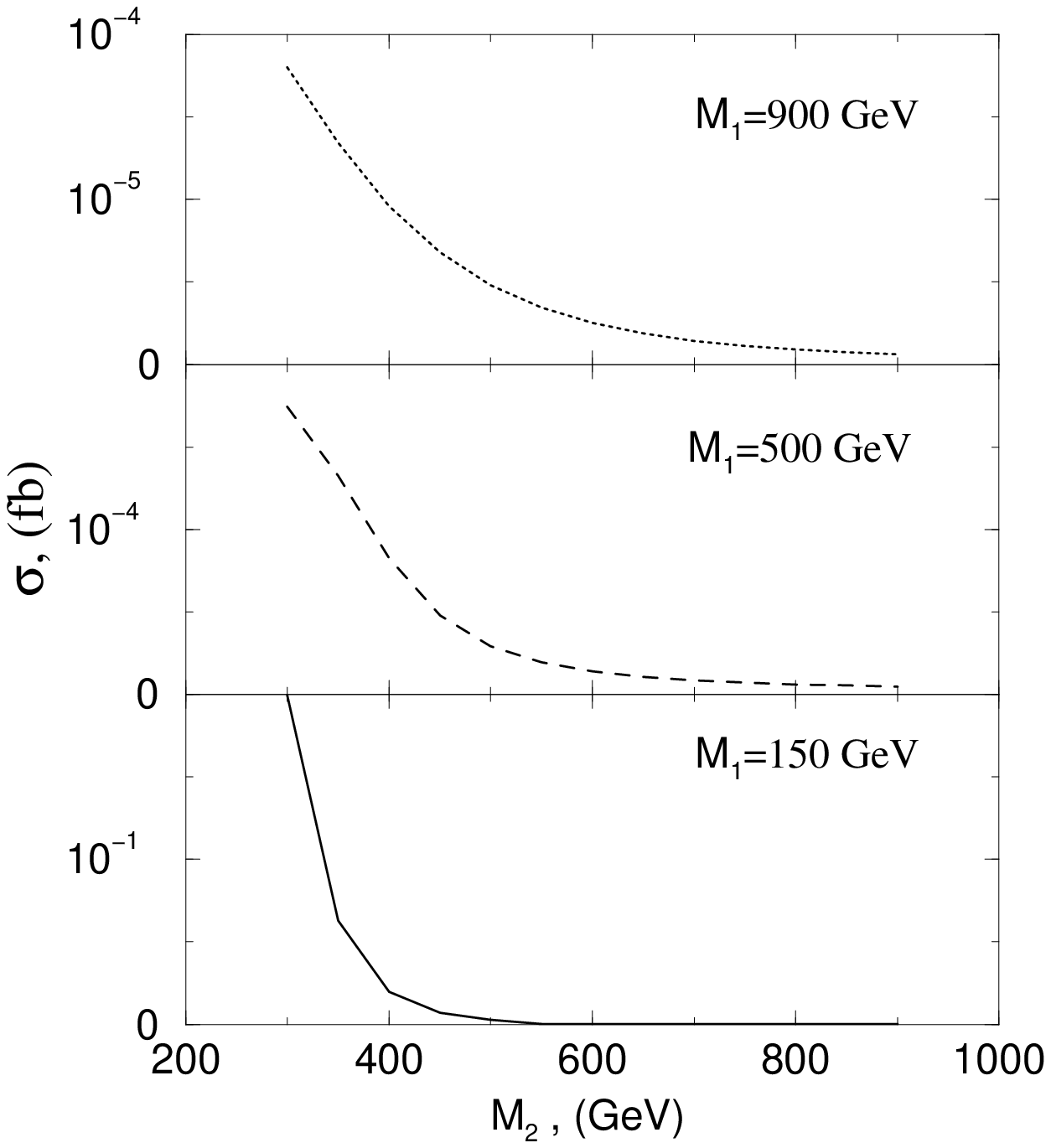
                             ,height=7.1cm,width=7.5cm}}
      \end{center}
\caption{The cross sections of the subprocess
$u\bar{u}\to\widetilde{\chi}_{2}^{0}\widetilde{\chi}_{2}^0$, as a
function of $M_2$ with $\mu = 450 $ GeV and $\sqrt{\hat{s}} =1.5$
TeV}
\label{Fig10}
\end{figure}

\begin{figure}[hpb]
    \begin{center}
\mbox{\epsfig{figure=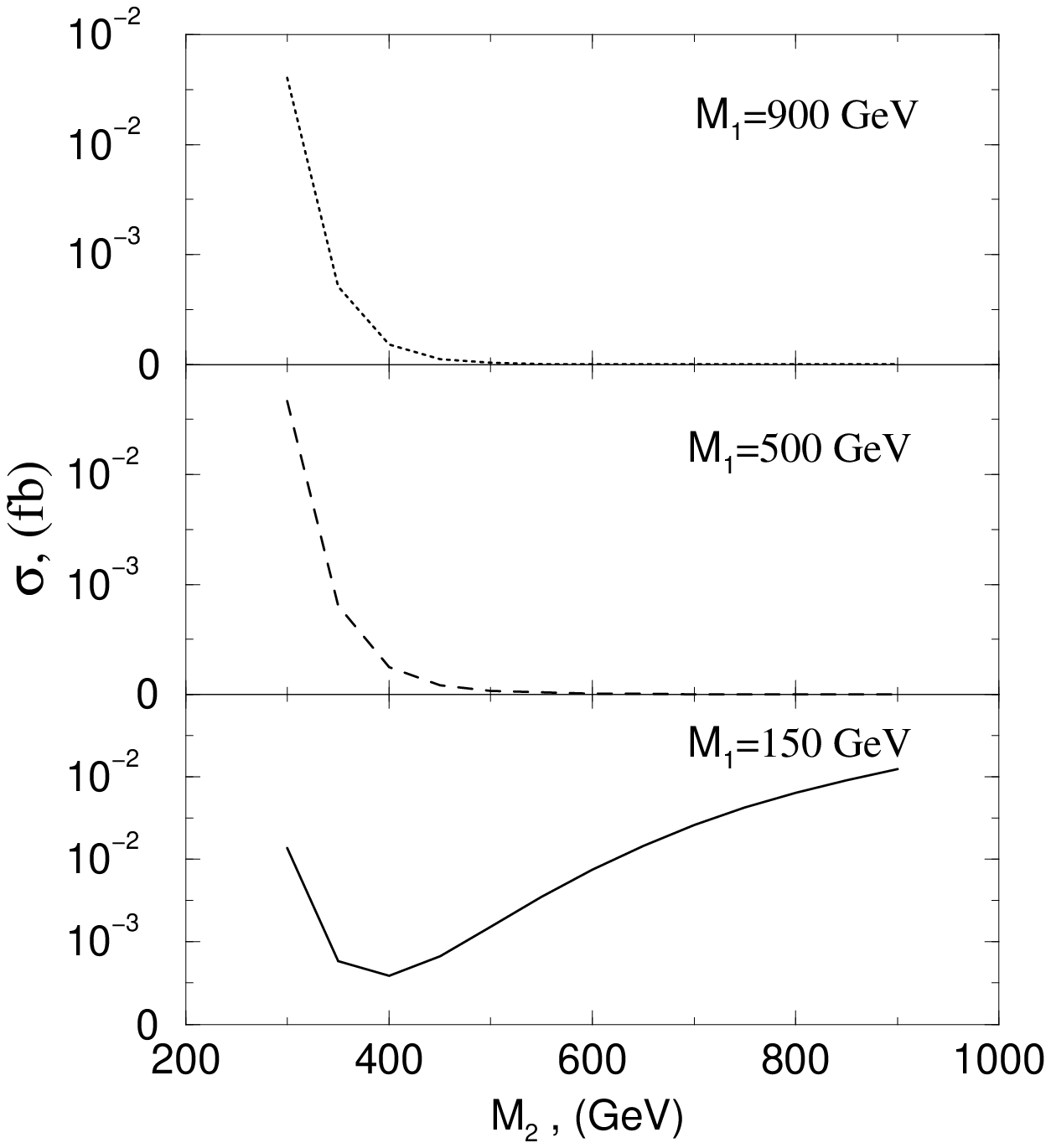
                             ,height=6.9cm,width=7.7cm}}
      \end{center}
\caption{The cross sections of the subprocess
$d\bar{d}\to\widetilde{\chi}_{1}^{0}\widetilde{\chi}_{1}^0$, as a
function of $M_2$ with $\mu = 450 $ GeV and $\sqrt{\hat{s}} =1.5$
TeV}
\label{Fig11}
\end{figure}

\newpage

\begin{figure}[hpb]
    \begin{center}
\mbox{\epsfig{figure=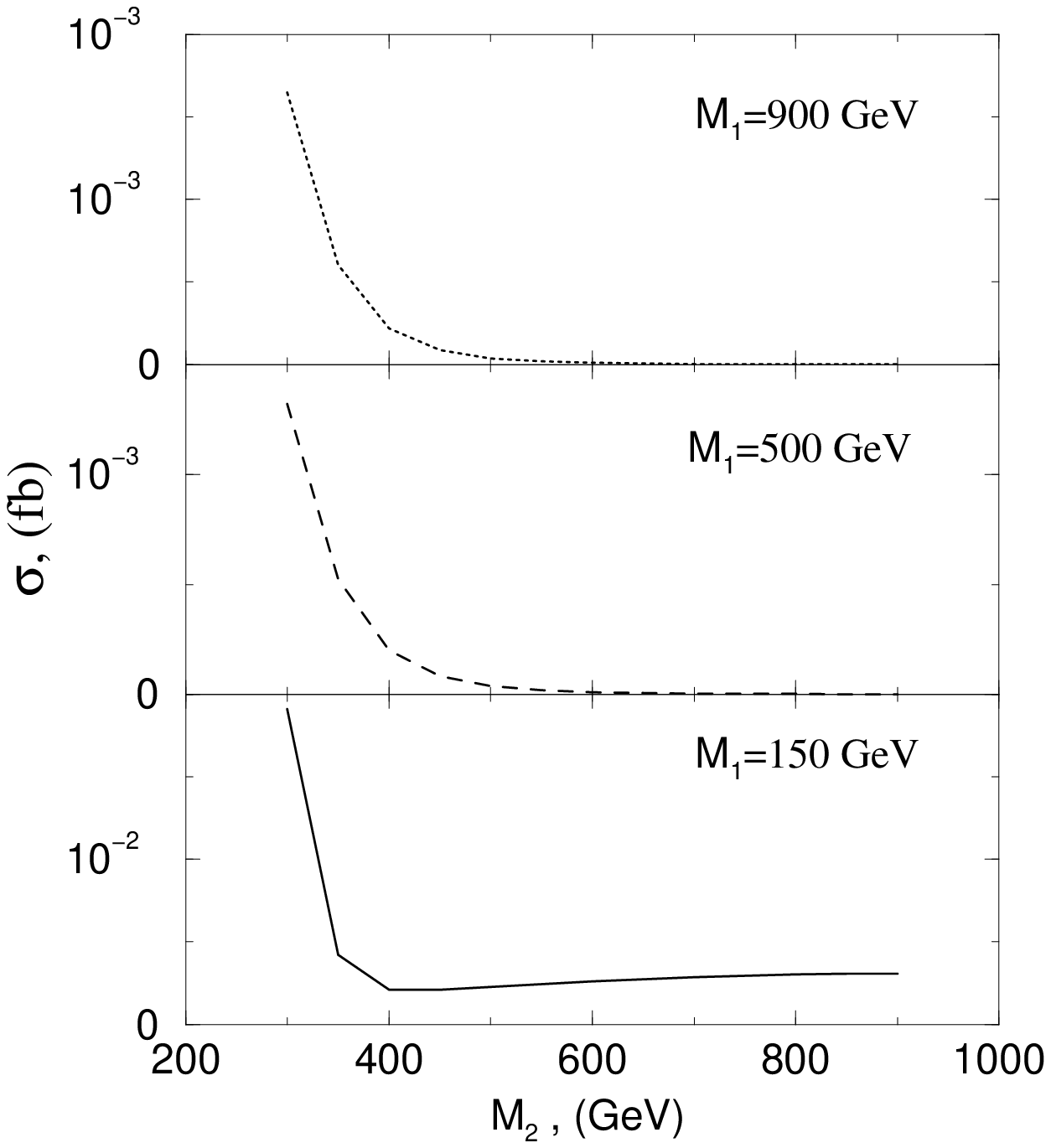
                             ,height=6.9cm,width=7.7cm}}
      \end{center}
\caption{The cross sections of the subprocess
$d\bar{d}\to\widetilde{\chi}_{1}^{0}\widetilde{\chi}_{2}^0$, as a
function of $M_2$ with $\mu = 450 $ GeV and $\sqrt{\hat{s}} =1.5$
TeV}
\label{Fig12}
\end{figure}

\begin{figure}[hpb]
    \begin{center}
\mbox{\epsfig{figure=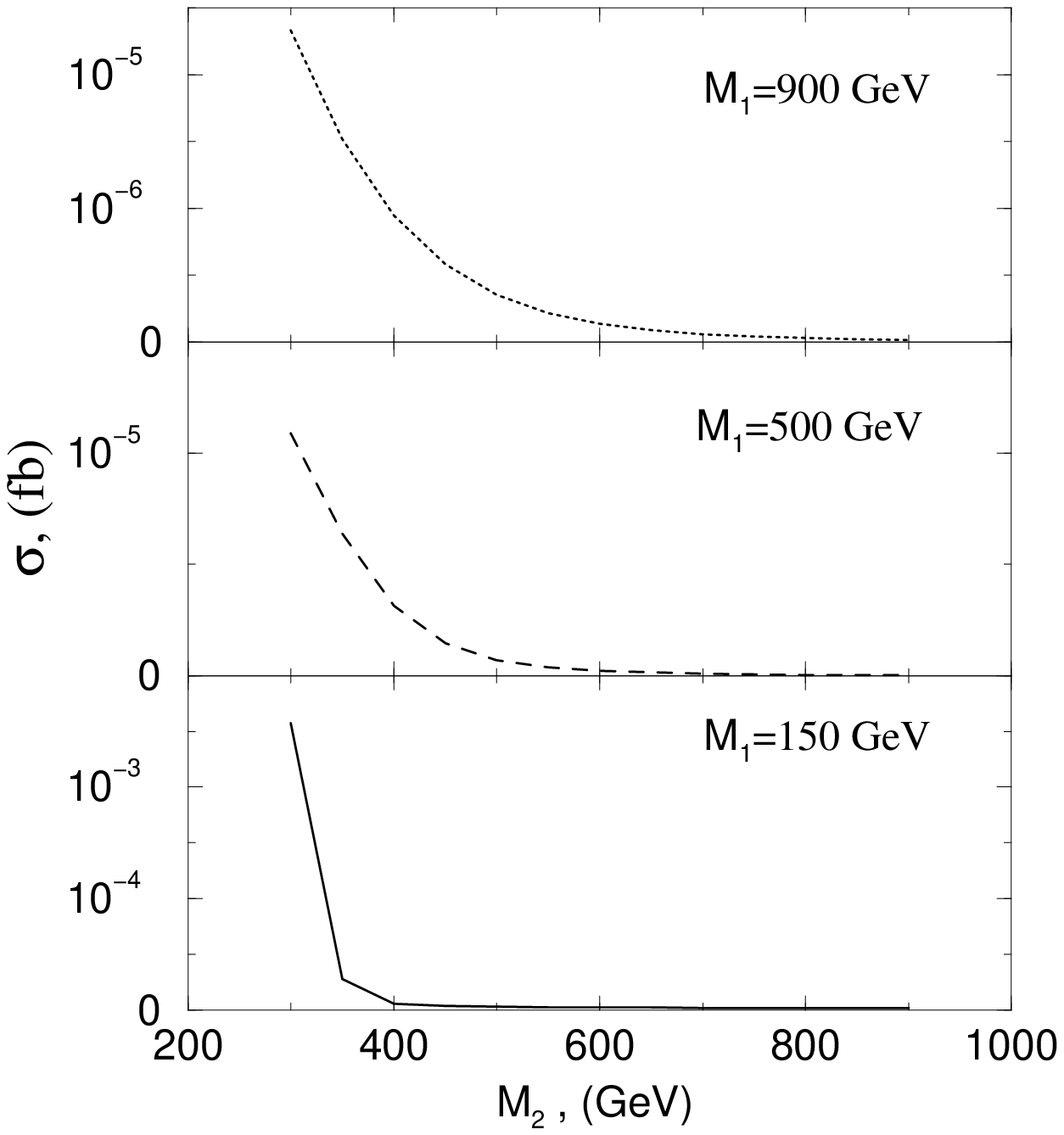
                             ,height=6.7cm,width=7.5cm}}
      \end{center}
\caption{The cross sections of the subprocess
$d\bar{d}\to\widetilde{\chi}_{2}^{0}\widetilde{\chi}_{2}^0$, as a
function of $M_2$ with $\mu = 450 $ GeV and $\sqrt{\hat{s}} =1.5$
TeV}
\label{Fig13}
\end{figure}

\newpage

\begin{figure}[hpb]
    \begin{center}
\mbox{\epsfig{figure=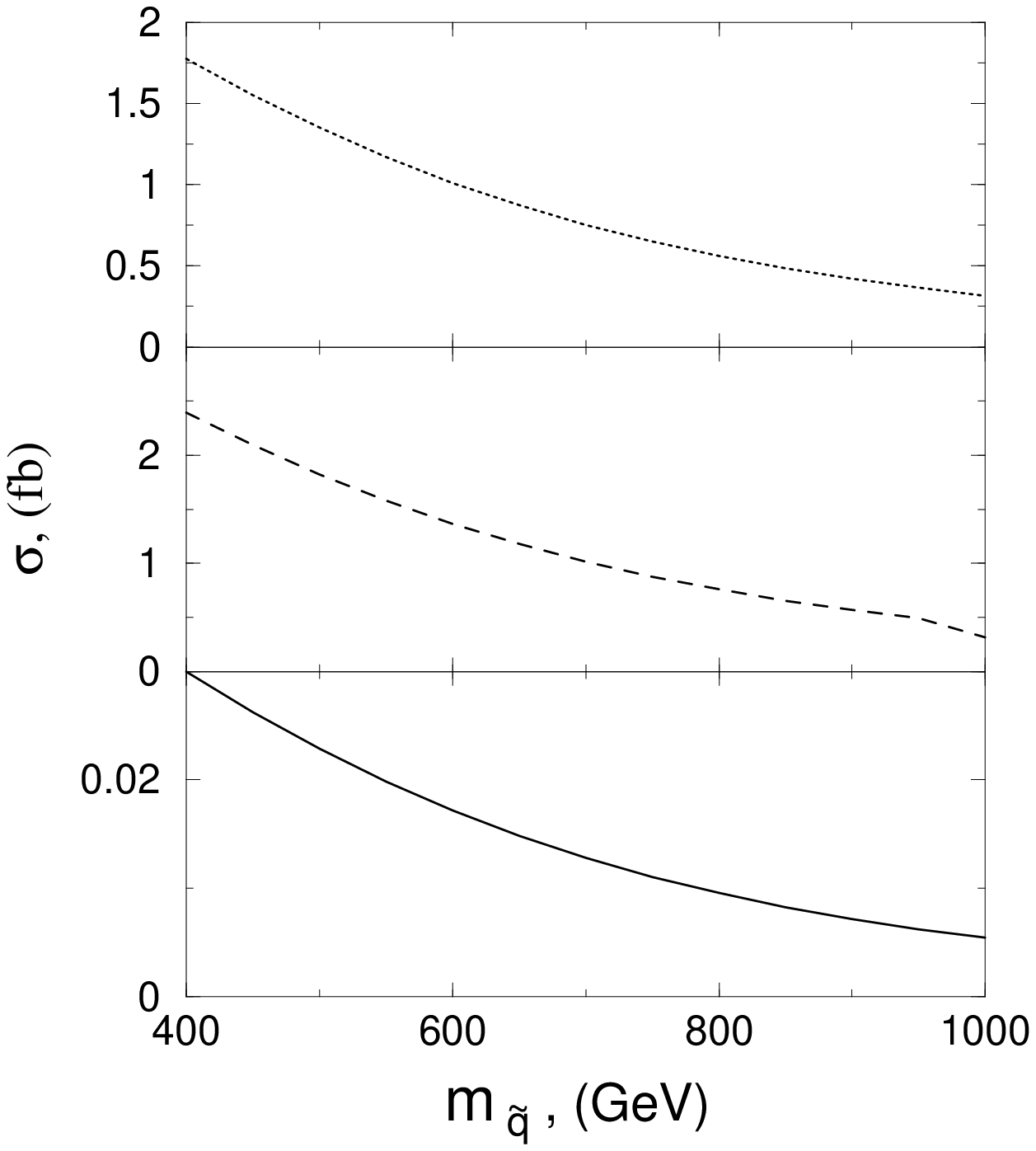
                             ,height=6.7cm,width=7.5cm}}
      \end{center}
\caption{The cross sections of the subprocess
$u\bar{u}\to\widetilde{\chi}_{1}^{0}\widetilde{\chi}_{1}^0$, as a
function of the squark mass at beam energy $\sqrt{\hat{s}} =1.5$
TeV The curves correspond to: solid-Higgsino-like,
dashed-gaugino-like and dotted-mixture cases, respectively.}
\label{Fig14}
    \begin{center}
\mbox{\epsfig{figure=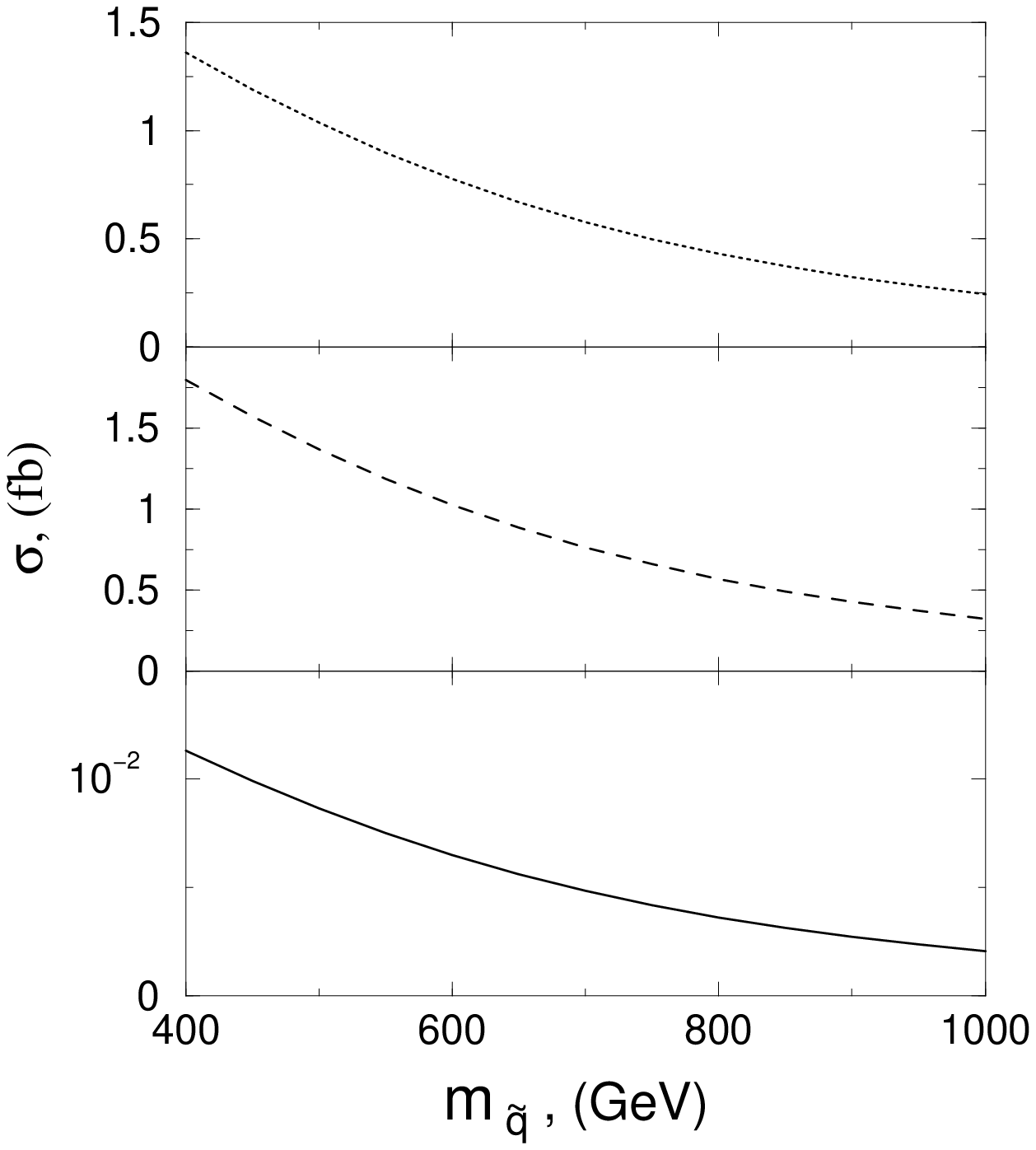
                             ,height=6.5cm,width=7.5cm}}
      \end{center}
\caption{The cross sections of the subprocess
$u\bar{u}\to\widetilde{\chi}_{1}^{0}\widetilde{\chi}_{2}^0$, as a
function of the squark mass at beam energy $\sqrt{\hat{s}} =1.5$
TeV The curves correspond to: solid-Higgsino-like,
dashed-gaugino-like, and dotted-mixture cases, respectively.}
\label{Fig15}
\end{figure}

\newpage

\begin{figure}[hpb]
    \begin{center}
\mbox{\epsfig{figure=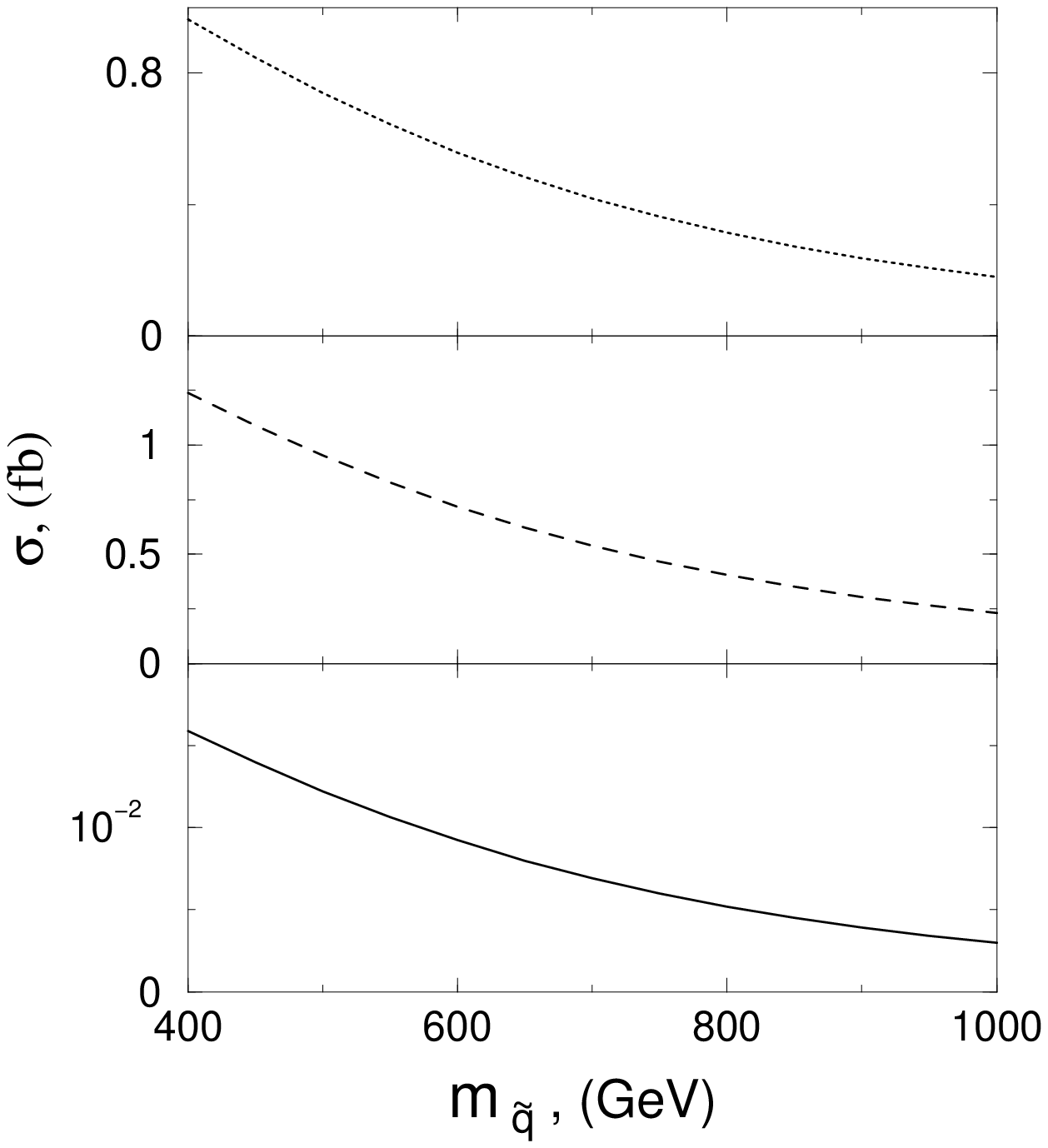
                             ,height=6.5cm,width=7.5cm}}
      \end{center}
\caption{The cross sections of the subprocess
$u\bar{u}\to\widetilde{\chi}_{2}^{0}\widetilde{\chi}_{2}^0$, as a
function of the squark mass at beam energy $\sqrt{\hat{s}} =1.5$
TeV The curves correspond to: solid-Higgsino-like,
dashed-gaugino-like, and dotted-mixture cases, respectively.}
\label{Fig16}
\end{figure}

\begin{figure}[hpb]
    \begin{center}
\mbox{\epsfig{figure=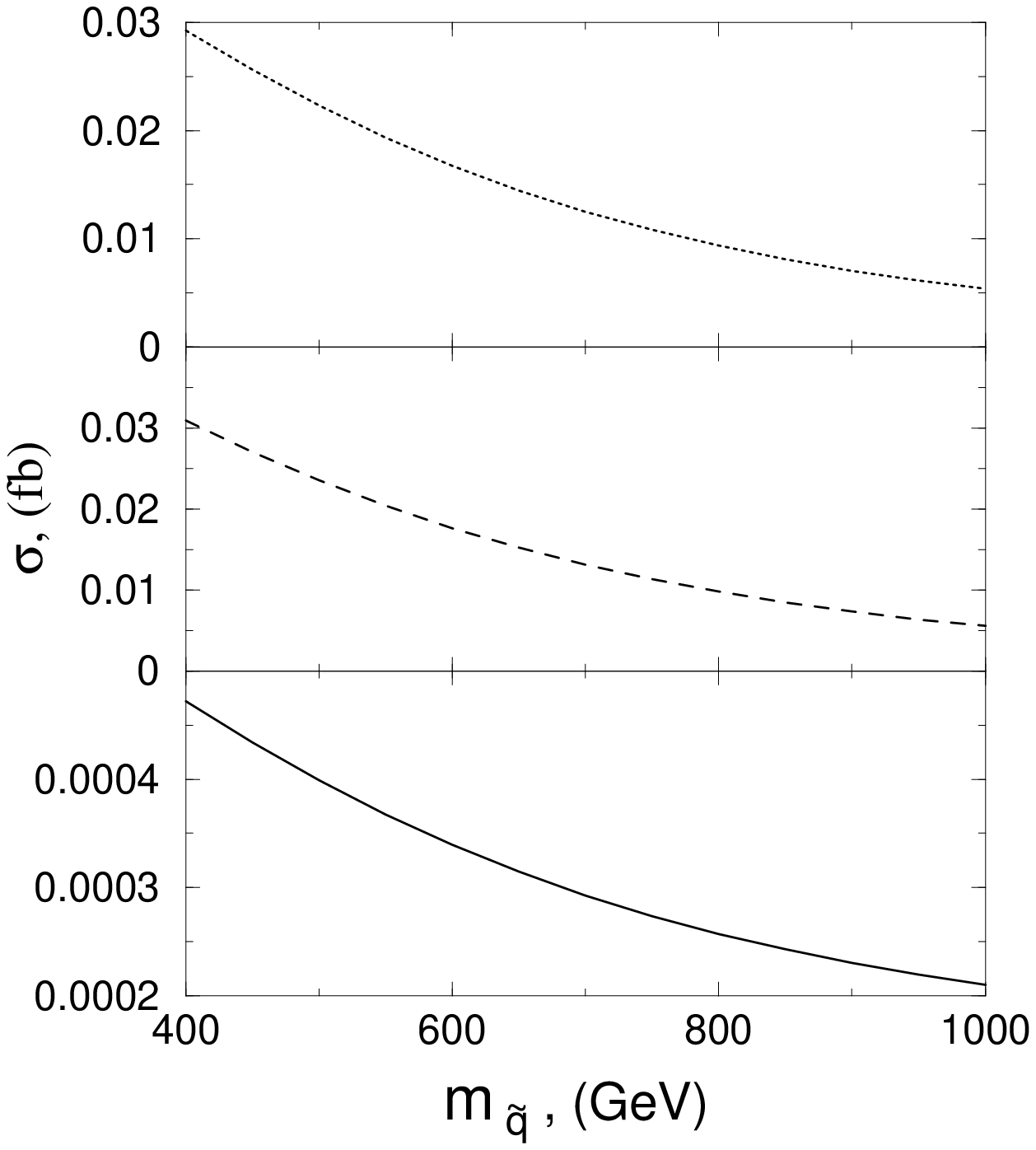
                             ,height=6.5cm,width=7.5cm}}
      \end{center}
\caption{The cross sections of the subprocess
$d\bar{d}\to\widetilde{\chi}_{1}^{0}\widetilde{\chi}_{1}^0$, as a
function of the squark mass at beam energy $\sqrt{\hat{s}} =1.5$
TeV The curves correspond to: solid-Higgsino-like,
dashed-gaugino-like and dotted-mixture cases, respectively.}
\label{Fig17}
\end{figure}

\newpage

\begin{figure}[hpb]
    \begin{center}
\mbox{\epsfig{figure=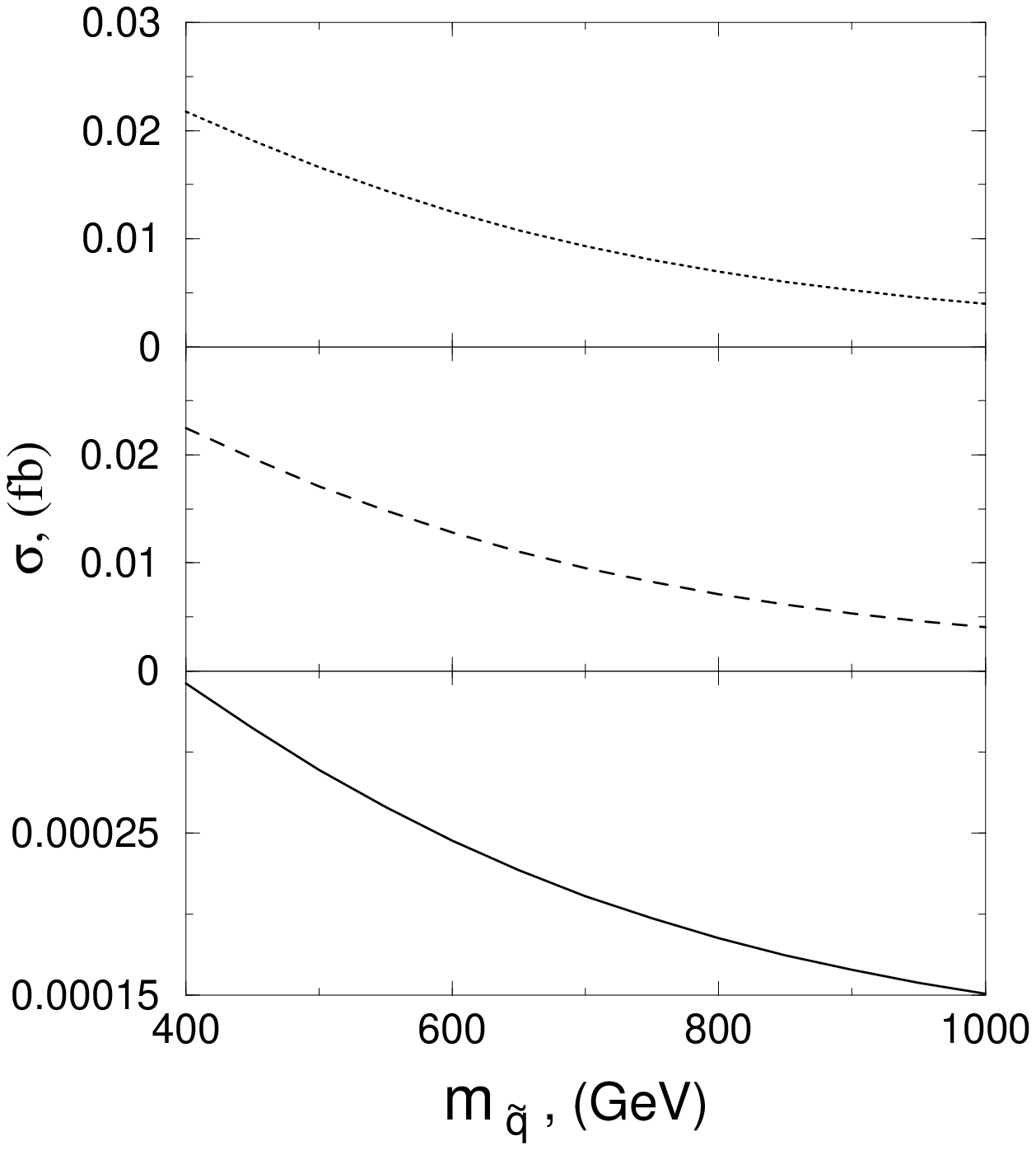
                             ,height=6.5cm,width=7.5cm}}
      \end{center}
\caption{The cross sections of the subprocess
$d\bar{d}\to\widetilde{\chi}_{1}^{0}\widetilde{\chi}_{2}^0$, as a
function of the squark mass at beam energy $\sqrt{\hat{s}} =1.5$
TeV The curves correspond to: solid-Higgsino-like,
dashed-gaugino-like, and dotted-mixture cases respectively.}
\label{Fig18}
\end{figure}

\begin{figure}[hpb]
    \begin{center}
\mbox{\epsfig{figure=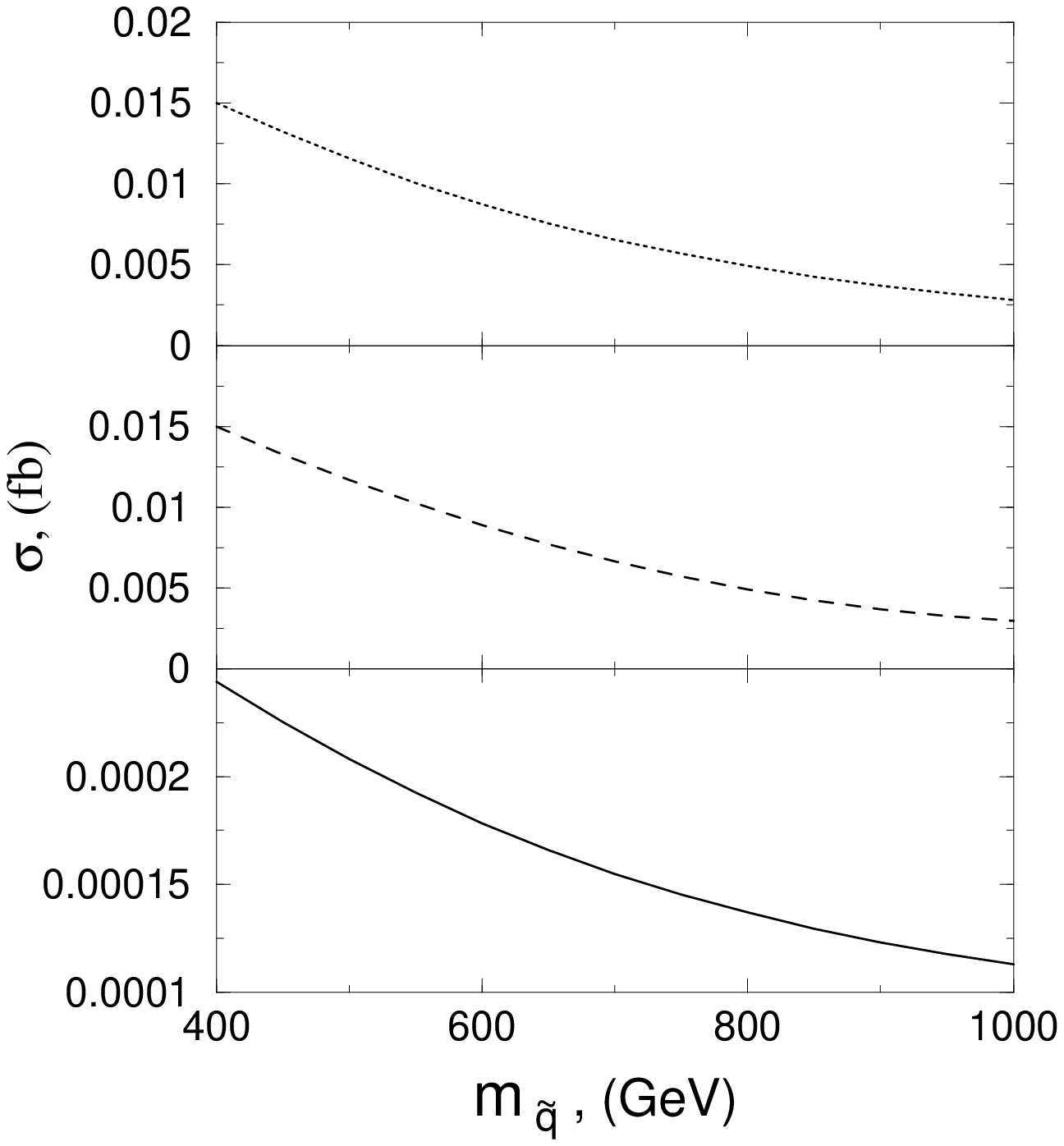
                             ,height=6.5cm,width=7.5cm}}
      \end{center}
\caption{The cross sections of the subprocess
$d\bar{d}\to\widetilde{\chi}_{2}^{0}\widetilde{\chi}_{2}^0$, as a
function of the squark mass at beam energy $\sqrt{\hat{s}} =1.5$
TeV. The curves correspond to: solid-Higgsino-like,
dashed-gaugino-like and dotted-mixture cases, respectively.}
\label{Fig19}
\end{figure}

\newpage

\begin{figure}[hpb]
    \begin{center}
\mbox{\epsfig{figure=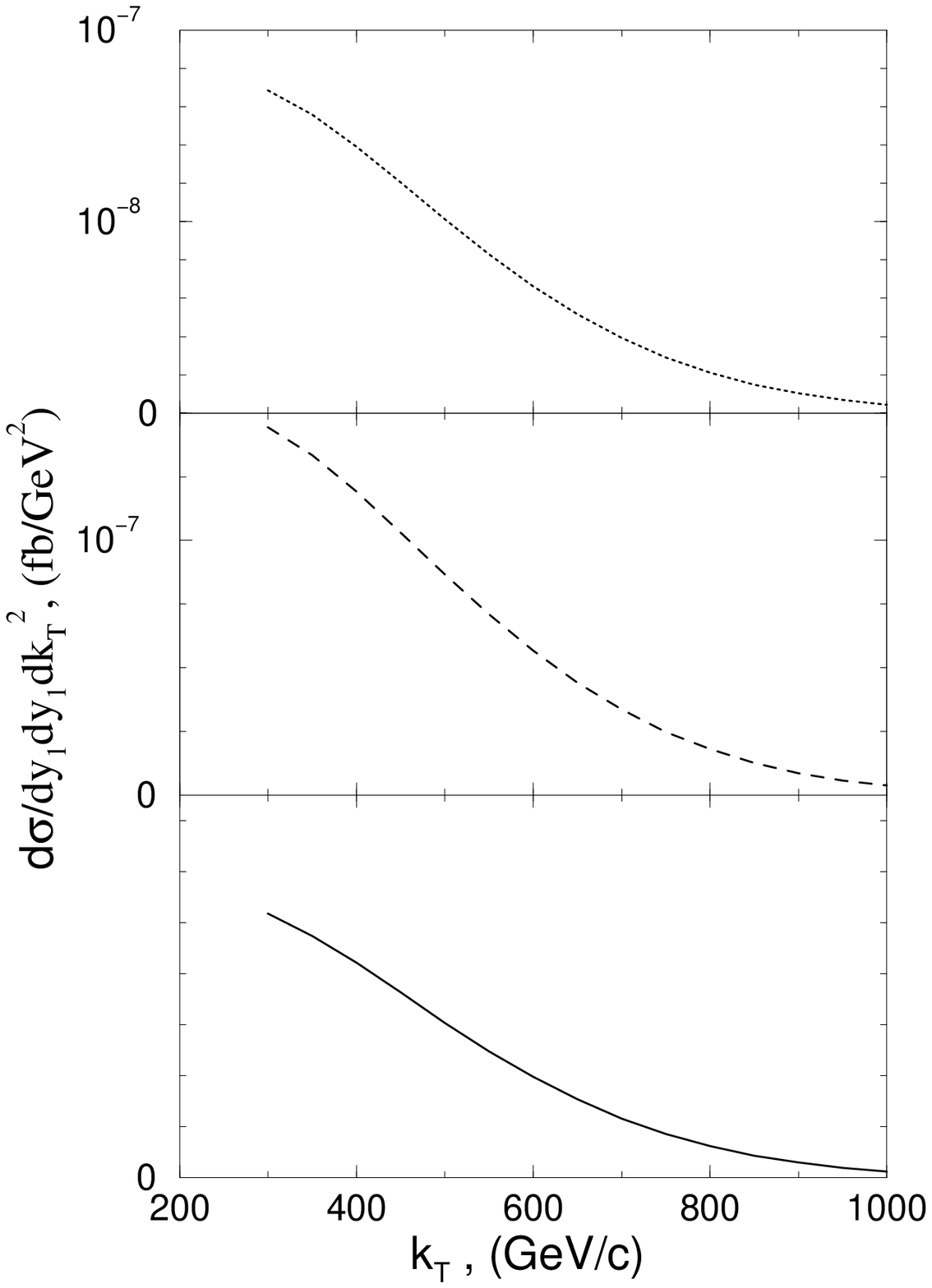
                             ,height=6.5cm,width=7.5cm}}
      \end{center}
\caption{The differential cross sections of the process $pp \to
\widetilde{\chi}_{1}^0\widetilde{\chi}_{1}^0$ as a function of the
$k_T$ transverse momentum of the neutralino pair. The curves
correspond to: solid-Higgsino-like, dashed-gaugino-like and
dotted-mixture cases, respectively.}
\label{Fig20}
\end{figure}

\begin{figure}[hpb]
    \begin{center}
\mbox{\epsfig{figure=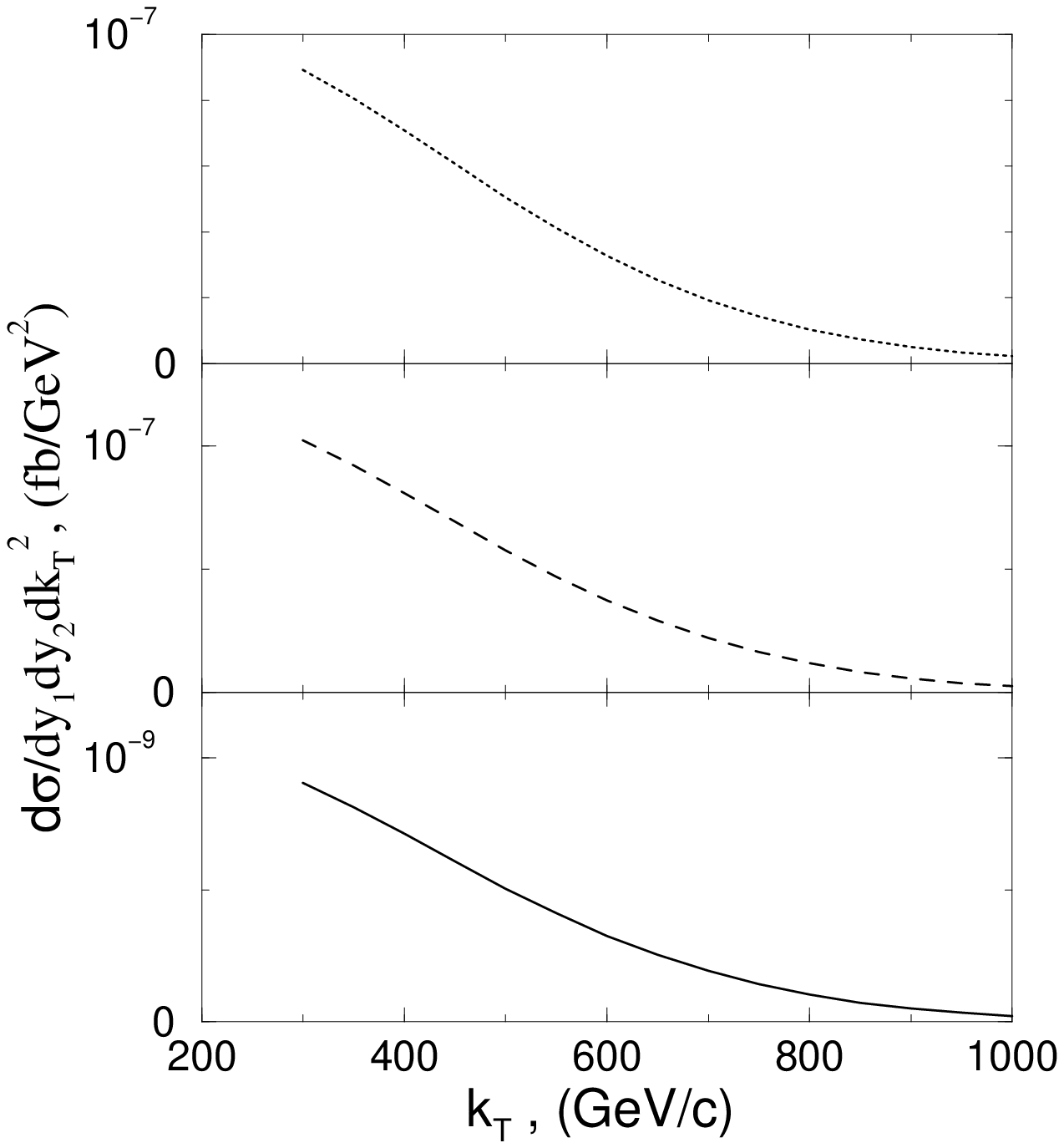
                             ,height=6.5cm,width=7.5cm}}
    \end{center}
\caption{The differential cross sections of the process $pp \to
\widetilde{\chi}_{1}^0\widetilde{\chi}_{2}^0$ as a function of the
$k_T$ transverse momentum of the neutralino pair. The curves
correspond to: solid-Higgsino-like, dashed-gaugino-like and
dotted-mixture cases, respectively.}
\label{Fig21}
\end{figure}

\newpage

\begin{figure}[hpb]
     \begin{center}
\mbox{\epsfig{figure=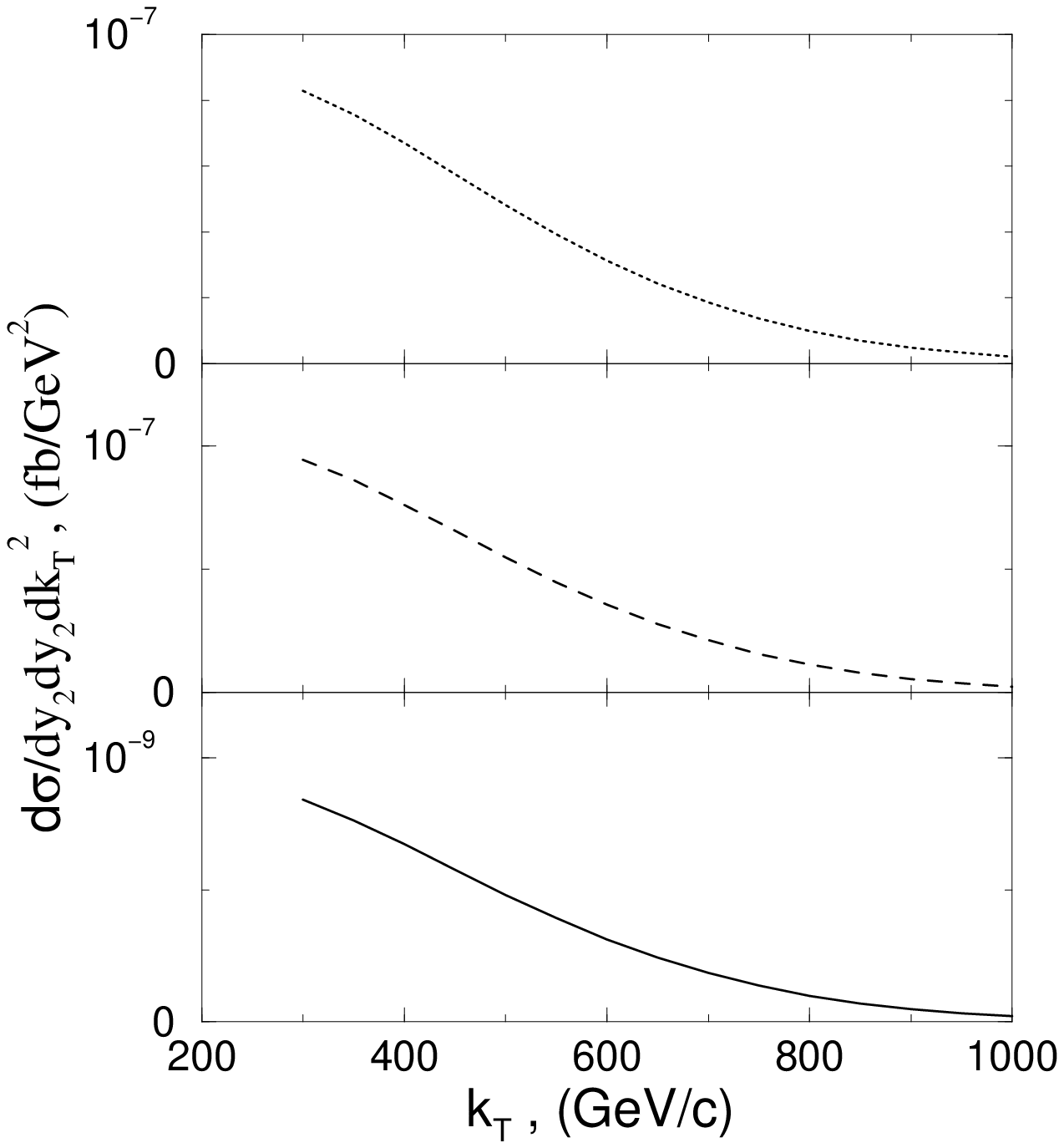
                            ,height=6.5cm,width=7.5cm}}
      \end{center}
\caption{The differential cross sections of the process $pp \to
\widetilde{\chi}_{2}^0\widetilde{\chi}_{2}^0$ as a function of the
$k_T$ transverse momentum of the neutralino pair. The curves
correspond to: solid-Higgsino-like, dashed-gaugino-like and
dotted-mixture cases, respectively.} \label{Fig22}
\end{figure}

\end{document}